\documentclass[preprint,amsmath,amssymb,nofootinbib]{revtex4}

\usepackage{color}
\usepackage{graphicx}

\usepackage{dcolumn}
\usepackage{bm}
\usepackage{lscape}
\usepackage{ulem}

\usepackage{txfonts}

\begin{document}
\title{Systematic study for relation between nuclear structure and reaction in $^{10}$Be nucleus}

\author{T. Furumoto}%
\email{furumoto-takenori-py@ynu.ac.jp}
\affiliation{College of Education, Yokohama National University, Yokohama 240-8501, Japan}

\author{T. Suhara}%
\affiliation{Matsue College of Technology, Matsue 690-8518, Japan}

\author{N. Itagaki}%
\affiliation{Department of Physics, Osaka Metropolitan University, Osaka, 558-8585, Japan}
\affiliation{Nambu Yoichiro Institute of Theoretical and Experimental Physics (NITEP), Osaka Metropolitan University, Osaka 558-8585, Japan}

\date{\today}

\begin{abstract}
We systematically investigate the relation between the nuclear structure and reaction in the $^{10}$Be nucleus using a theoretical framework.
The structure of the $^{10}$Be nucleus is constructed with a cluster model based on a microscopic viewpoint.
In this paper, the $^{10}$Be nucleus with different structures is prepared by manipulating the parameters of an effective nucleon-nucleon interaction.
The nuclear structure and expectation values of physical quantities are drastically changed by the modification.
We summarize such changes and show the effects on the elastic and inelastic scatterings for the proton and $^{12}$C targets in the microscopic coupled-channel calculation.
Especially, we recently reported the visualization of dineutron correlation in $^{10}$Be on proton inelastic scattering in [Phys. Rev. C104, 034613 (2021)].
In this preceding work, we found that the changing the degree of dineutron correlation in $^{10}$Be leads to drastic changes of the inelastic cross section for the 2$_2^+$ state.
The development (or breaking) of the dineutron correlation is
governed by the strength of the spin-orbit interaction of the structure calculation.
However, in the previous work, some of the realistic physical points were missing, for example, the binding energy.
Therefore, we reconstruct the $^{10}$Be nucleus by adjusting the effective nucleon-nucleon interaction to obtain the reasonable binding energy of 
the ground state.
With this improvement, we again discuss the dineutron correlation in the $^{10}$Be nucleus.
We reconfirm the way to measure the degree of the development (or breaking) of dineutron cluster structure; the sensitivity to the inelastic cross section of the ground state to the 2$_2^+$ state of $^{10}$Be.
\end{abstract}

\maketitle

\section{Introduction}
For decades, the microscopic description based on the nucleon-nucleon ($NN$) interaction for the nuclear structure and reaction has been developed.
These microscopic approaches are successful in describing and understanding the properties of the nuclear structure and reaction quite well.
Although the $NN$ interaction has been investigated since the discovery of the proton and neutron, its behavior is still not fully understood, especially in the nuclear medium.
Therefore, in these microscopic structure and reaction calculations,
the effective $NN$ interaction with parameters has often been used.
There are many microscopic models based on the (effective) $NN$ interaction for the nuclear structure~\cite{VAU72, DEC80, KAM81, ONO92, BAY94, ITA95, NAV00, OTS01, NAK03, SUH10, ISA15, Tohsaki:2001an, Kanada-Enyo:1995jpd, Neff:2003ib}.
The microscopic description has led to the understanding of a great deal of nuclear structure information.
Recently, improvements in computational technology have made it possible to perform even more precise nuclear structure calculations.
Most microscopic approaches to nuclear reactions are based on the folding model.
This has been developed for the nucleon and heavy-ion scatterings~\cite{SIN75,BRI77,BRI77-2,BRI78,SAT79,KOB82,KHO94, RIK842,KHO97,MEL00,KHO01,FUR06,KHO07,CEG07,FUR09,FUR10,FUR16}.
It is remarkable that the recent microscopic description of the nuclear structure and reaction is based on the realistic $NN$ interaction.

Through these analyses, the connection between the microscopic nuclear structure and reaction models has been attempted.
By the connection, microscopic nuclear reaction calculations accurately incorporating nuclear structure information were performed, and many experimental data were successfully reproduced~\cite{SAK86, TAK05, KHO11, WWQ15, WWQ17, KAN19}.
In addition, the approach with the microscopic nuclear structure and reaction gives interesting predictions~\cite{FUR10, FUR13-1, FUR13-2, FUR21-2}.
In such approaches, density is one of the key intermediaries because it is one of the inputs in the microscopic folding model and one of the outputs in the microscopic nuclear structure model.
The diagonal and transition densities contain a large amount of nuclear structure information.
The microscopic folding model tells us the nuclear structure information in the observable through the density.
Therefore, to derive the diagonal and transition densities not only microscopically but also macroscopically is important to connect nuclear structure and reaction.
Here, we note that it is possible to observe the structure information in experimental data if the diagonal and transition densities have a sensitivity to the nuclear structure.
Therefore, it is important to make the relation among density, nuclear structure, reaction mechanism, and observable clearer.

Here, we demonstrate an example of how the change of the nuclear structure in an artificial way affects the reaction.
The change is closely related to the development and breaking of the dineutron correlation as reported in Ref.~\cite{ITA02}, where we discussed the dineutron correlation and its breaking because of the spin-orbit interaction.
The persistence of the dineutron cluster is sensitive to the choice of the strength of the spin-orbit interaction in the structure calculation, even if the binding energy of the neutrons from the threshold is kept constant.
Although the dineutron structure is favored when the spin-orbit interaction is weak, the spin-orbit interaction with realistic strength significantly breaks the dineutron structure.
In Ref.~\cite{FUR21-2}, we found that the inelastic cross section to the 2$^+_2$ state is drastically dependent on the development and breaking of the dineutron correlation.
With the advancement of microscopic approaches, such a reliable and interesting analysis is now feasible.
Here, we again summarize the basic properties of $^{10}$Be.
The $^{10}$Be nucleus is considered as a typical neutron-rich nucleus with a cluster structure, and in this nucleus, two valence neutrons perform molecular-orbital motions around two $\alpha$ clusters~\cite{SEY81, ITA00-1, ITA00-2, ITA02}.
With regard to its 0$^+$ states, the 0$^+_1$ and 0$^+_3$ states are characterized by the $\pi$ orbit of the valence neutrons around $\alpha$--$\alpha$.
On the other hand, the 0$^+_2$ state has a large $\alpha$--$\alpha$ distance, which is characterized by the $\sigma$ orbit.
Above these 0$^+$ states, rotational bands are formed.
For example, the 2$^+_1$ state is a member of the $K=0$ band together with the ground $0^+$ state, and the 2$^+_2$ state is the band head state of the $K=2$ side-band.

In this study, we perform the systematic calculation to reveal the relation between the nuclear structure and reaction of the $^{10}$Be nucleus.
The present $^{10}$Be nucleus is constructed under the assumption of the four-body ($\alpha$ + $\alpha$ + n + n) cluster model.
The stochastic multi-configuration mixing method enables to the description of many exotic cluster structures~\cite{ICH11, MUT11}.
We focus on the low-lying (0$_1^+$, 2$_1^+$, and 2$_2^+$) states.
In the microscopic cluster model, we modify the various parameters for the effective $NN$ interaction.
The energies, nuclear size, and expectation values of $\langle \bm{L} \cdot \bm{S} \rangle$ (one-body operator) and $\langle \bm{S}^2 \rangle$ (two-body operator) are calculated to investigate the dependence of the parameters for the $NN$ interaction.
The transition strengths with multipolarity $\lambda = 2$ are listed for the proton, neutron, isoscalar (IS), and isovector (IV) parts, respectively.
To apply the microscopic nuclear reaction model, we obtain the diagonal and transition densities from the total wave function.
Specifically, the microscopic coupled channel (MCC) calculation is performed to describe $^{10}$Be elastic and inelastic scatterings by a proton and $^{12}$C targets at $E/A =$ 59.4 and 200 MeV.
The present MCC calculation well reproduces the experimental data for the 0$_1^+$ and 2$_1^+$ states at $E/A =$ 59.4 MeV.
Although no experimental data are available for the 2$_2^+$ state, the inelastic cross section is demonstrated in the present MCC calculation.
The effect of the change of the parameters in the microscopic structure calculation is discussed on the elastic and inelastic scattering cross sections derived from the MCC calculation.
In addition, we reconstruct the $^{10}$Be nucleus by adjusting the binding energy, which was ignored in the previous systematic calculation.
With this improvement, we again discuss the possibility of observing the degree of the development and breaking of the dineutron correlation in $^{10}$Be through the 2$_2^+$ inelastic cross section.

This paper is organized as follows.
In Sec.~\ref{sec:formalism}, we briefly introduce the present structure and reaction models.
In Sec.~\ref{sec:results}, we listed the results of the microscopic cluster model with the modified parameters.
We also show the results of the elastic and inelastic cross sections with the modified nuclear structure information.
We will discuss the effect of the value of the parameters in the nuclear structure calculation.
In addition, we show the results of the reconstructed $^{10}$Be nucleus by reproducing the binding energy by simultaneously adjusting the spin-orbit strength and the Majorana parameter.
The development and breaking of the dineutron correlation in $^{10}$Be are again discussed.
Lastly, we summarize this paper in Sec.~\ref{sec:summary}.

\section{Formalism}
\label{sec:formalism}

We first construct the $^{10}$Be nucleus within the 4-body ($\alpha$ + $\alpha$ + n + n) cluster model.
With the diagonal and transition densities obtained from the microscopic cluster calculation, the $^{10}$Be + p and $^{10}$Be + $^{12}$C elastic and inelastic scatterings are given by the microscopic coupled channel (MCC) calculation as the same manners in Refs.~\cite{FUR13-2, FUR21-2}.
To avoid repetition, we give just the main points in this section.
The details of the structure and reaction calculations are provided in Refs.~\cite{FUR13-2, FUR18, FUR21-1, FUR21-2}.

\subsection{Microscopic cluster model}
The $^{10}$Be nucleus is constructed by the stochastic multiconfiguration mixing method based on the microscopic cluster model~\cite{ICH11, MUT11}.
The calculation method is almost the same as Ref.~\cite{FUR21-2}.
Therefore, we use a simplified notation in this paper. 

The total wave function $\Phi$ is expressed by the superposition of the basis states $\Psi$ as follows:
\begin{equation}
\Phi = \sum_{i} c_{i} \Psi_{i}.
\label{mcmwf}
\end{equation}
After the parity and angular momentum projections, the eigenstates of the Hamiltonian are obtained by diagonalizing the Hamiltonian matrix.
Additionally, the coefficients, $c_{i}$, for the linear combination of Slater determinants are obtained.
We adopt 500 basis states to obtain the total wave function.
We confirmed that sufficient convergence was achieved with 500 basis states.
In fact, for the 4-body cluster system, even 400 basis states are known to provide sufficient convergence as shown in Ref.~\cite{FUR13-2}.

In the above process, we introduce various $\alpha$ + $\alpha$ + n + n configurations for the basis states to describe the $^{10}$Be nucleus as follows:
\begin{equation}
\Psi_{i} = {\cal A}
 \big[ \phi_{\alpha} (\bm{r}_{1-4}, \bm{R}_{1}) \phi_{\alpha} (\bm{r}_{5-8}, \bm{R}_{2}) 
\phi_{n} (\bm{r}_{9}, \bm{R}_{3}) \phi_{n} (\bm{r}_{10}, \bm{R}_{4}) \big]_{i},
\end{equation}
where $\cal A$ is the antisymmetrizer and $\phi_{\alpha}$ and $\phi_{n}$ are the wave functions of $\alpha$ and neutron, respectively.
The wave functions of the $j$-th nucleon, whose spatial coordinate is $\bm{r}_{j}$, is described as a locally shifted Gaussian centered at $\bm{R}$, $\exp[-\nu(\bm{r}_j - \bm{R})^2]$.
Here, the positions of the Gaussian-centered parameter $\bm{R}$ are randomly generated.
The $\nu$ parameter is $\nu=$ 1/(2$\times$1.46$^2$) fm$^{-2}$, which is widely used and satisfies the binding energy of the $\alpha$ particle.

The Hamiltonian is given as follows.
The two-body interaction includes the central, spin-orbit, and Coulomb parts.
The Volkov No.2 effective potential is applied to the central part~\cite{VOL65}, as,
\begin{eqnarray}
V^{\rm (central)}(r) &=&( W-MP^{\sigma}P^{\tau}+ BP^{\sigma}-HP^{\tau})\nonumber \\
&& \times (V_{1}\exp(-r^{2}/ c_{1}^{2})+ V_{2}\exp(-r^{2}/ c_{2}^{2})),
\end{eqnarray}
 where $c_{1}=$ 1.01 fm, $c_{2}= 1. 8$ fm, $V_{1}=$ 61.14 MeV, $V_{2}=-60. 65$ MeV, and $W= 1-M$.
Here, we often use the parameter setting as $M = 0.60$ and $B = H = 0.08$~\cite{FUR13-2, FUR18, FUR21-2}, which reproduces the $\alpha$-$\alpha$ scattering phase shift.
Meanwhile, the different values for the parameters $M$ and $B=H$ are employed in the investigations of various nuclei~\cite{VAR93, ITO04, KAN07, SUH10}.
Therefore, we examine in the range of 0.52, 0.54, 0.56, 0.58, 0.60, 0.62, and 0.64 for $M$ and 0, 0.04, 0.08, 0.12, 0.16, and 0.20 for $B = H$.

Here, we introduce the spin-orbit term of the G3RS potential~\cite{TAM68, YAM79},
\begin{equation}
V^{\rm (spin-orbit)}(r) = V_{\rm LS}( e^{-d_{1}r^{2}}-e^{-d_{2}r^{2}}) P(^{3}O) \bm{L} \cdot {\bm{S}},
\end{equation}
where $d_{1}= 5. 0$ fm$^{-2}$ and $d_{2}= 2. 778$ fm$^{-2}$.
The operator $\bm{L}$ represents the relative angular momentum, and $\bm{S}$ represents the spin ($\bm{S}_1+\bm{S}_2$).
$P(^{3}O)$ is the projection operator onto the triplet odd state.
As in Ref.~\cite{FUR21-2}, the strength of the spin-orbit interaction, $V_{\rm LS}$, is also treated as a parameter in this paper.
This value is often fixed around 2000 MeV to reproduce the data of the $^{10}$Be nucleus.
In this paper, we show the results with $V_{\rm LS}$ = 0, 500, 1000, 1500, 2000, 2500, 3000, 3500, and 4000 MeV.

To connect the nuclear structure and reaction calculations, we prepare the diagonal and transition densities in the same manner as in Ref.~\cite{BAY94}.

\subsection{MCC model}
After calculating the diagonal and transition densities, we perform the nuclear reaction calculation.
We apply the calculated densities to MCC calculations with the complex $G$-matrix interaction MPa~\cite{YAM14, YAM16}.
The MPa interaction has been successfully applied for nuclear reactions~\cite{FUR16, WWQ17, FUR19, FUR21-1}.
The detailed MCC calculation procedure for the folding potential is described in Refs.~\cite{KHO02, CEG07, FUR21-1, FUR21-2}.
Then, we briefly introduce the single and double-folding model calculations.

The single-folding model potential is simply described as 
\begin{equation}
U^{(SF)}_{\alpha \to \beta}(R; E/A) = \int{ \rho_{I_{\alpha} \to I_{\beta}}(r) v(\bm{s}, \rho; E/A) d\bm{r} }, \label{eq:pot-sfm}
\end{equation}
where $R$ is the radial distance between the incident $^{10}$Be nucleus and the target proton.
$\alpha$ and $\beta$ mean channel number of the initial and final states.
$I_{\alpha}$ and $I_{\beta}$ are spins of the initial and final states, respectively.
$E/A$ is the incident energy per nucleon.
$\rho_{I_{\alpha} \to I_{\beta}}$ is the diagonal ($\alpha=\beta$) and transition ($\alpha \ne \beta$) densities; $s$ is the radial distance between a nucleon in the projectile nucleus and the target proton, and $\bm{s} = \bm{r} - \bm{R}$.
We note that the description of Eq.~(\ref{eq:pot-sfm}) is simplified; in the actual calculation, the proton and neutron densities are separately folded with the proton-proton and proton-neutron interactions, respectively.
The Coulomb potential is also obtained by folding the nucleon-nucleon Coulomb interaction and proton density.
The knock-on exchange part and the spin-orbit part are obtained in the same manner as in Refs.~\cite{FUR21-1, FUR21-2}.

The double-folding model potential is also provided as
\begin{equation}
U^{(DF)}_{\alpha \to \beta}(R; E/A) = \int{ \rho_{I_{\alpha} \to I_{\beta}}(\bm{r}) \rho'_{I'_{\alpha} \to I'_{\beta}}(\bm{r}') v(\bm{s}, \rho; E/A) d\bm{r} d\bm{r}' }, \label{eq:pot-dfm}
\end{equation}
where $I'_{\alpha}$ and $I'_{\beta}$ are spins of initial and final states for the target nucleus, respectively.
$\rho'_{I'_{\alpha} \to I'_{\beta}}$ is the diagonal ($\alpha=\beta$) and transition ($\alpha \ne \beta$) densities for the target nucleus.
Here, $\bm{s} = \bm{r} - \bm{r}' - \bm{R}$.
The frozen density approximation is applied to $\rho$ in $v$.
The frozen density approximation is the standard prescription to describe the nucleus-nucleus system and its efficiency is verified in Refs.~\cite{FUR09, FUR16}.
The knock-on exchange part is obtained in the same manner as Refs.~\cite{FUR13-1, FUR13-2}.

When the single and double folding potentials are applied to the nuclear reaction, we modify the strength of the imaginary part of the potential.
Since the complex $G$-matrix is constructed with the infinite nuclear matter, the strength of the imaginary part is often adjusted for the finite nucleus because these level densities are quite different.
Therefore, we consider the incident-energy-dependent renormalization factor, $N_W = 0.5 + (E/A)/1000$~\cite{FUR19}, for the imaginary part of the folding model potential.
We note that any additional parameter is not needed to calculate the $^{10}$Be scatterings by the proton and $^{12}$C targets.
With the potentials, the scattering matrix is obtained from the folded potentials by solving the CC equation based on the Stormer method.
Relativistic kinematics is used in the calculation.
The cross sections are calculated with the scattering amplitude derived from the scattering matrix as shown in Ref.~\cite{DNR}.

\section{Results}
\label{sec:results}
The calculated results will be introduced and discussed in this section.
We first show the dependence of the parameters ($V_{\rm LS}$, $M$, and $B(=H)$) for the binding energy, radius, and expectation values of $\langle \bm{L} \cdot \bm{S} \rangle$ and $\langle \bm{S}^2 \rangle$ of the 0$^+_1$, 2$^+_1$, and 2$^+_2$ states, respectively.
Here, $\langle \bm{L} \cdot \bm{S} \rangle$ is the expectation value of the one-body operator and the sum of the spin-orbit operator of the nucleons, whereas $\langle \bm{S}^2 \rangle$ is the two-body operator, which is the square of the sum of the one-body spin operator over the nucleons.
In the investigation of the dependence of the parameters ($V_{\rm LS}$, $M$, and $B(=H)$), we regard the standard values of $V_{\rm LS} =$ 2000 MeV, $M=$ 0.60, and $B=H=$ 0.08 as criterion.
The quadrupole transition strength between each states is also introduced.
In addition, we discuss the effect of the change in the nuclear structure on the elastic and inelastic cross sections.

When we calculate the $^{10}$Be + $^{12}$C system, we apply the $^{12}$C density obtained by 3$\alpha$-RGM~\cite{KAM81} to the MCC calculation.
The 0$^+_1$, 2$^+_1$, and 0$^+_2$ states of the $^{12}$C nucleus are included because the excitation states are well known to have a strong channel coupling effect.

After that, we discuss the relation between the dineutron correlation and the inelastic cross section at the realistic physical point.
Concretely, we show the additional results in which the $M$ and $V_{\rm LS}$ parameters are simultaneously adjusted without changing the binding energy of the ground state as in Ref.~\cite{ITA02}.
In order to evaluate the effect of the dineutron correlation on the structure and reaction, we prepare the pure dineutron basis and compare with more realistic one.

\subsection{Systematic calculation with microscopic cluster model}

Here we show the dependence of the structure on the parameters ($V_{\rm LS}$, $M$, and $B(=H)$).
The parameters give various binding energies, radii, and expectation values of $\langle \bm{L} \cdot \bm{S} \rangle$ and $\langle \bm{S}^2 \rangle$ for the 0$^+_1$, 2$^+_1$, and 2$^+_2$ states.
This systematic analysis clarifies the role of each parameter ($V_{\rm LS}$, $M$, and $B$ $(=H)$) in $^{10}$Be.
In addition, the quadrupole transition strengths between each state are examined.
The transition strengths are also changed together with the property of the nuclear structure.
We will discuss the effect of the property of the nuclear structure on the elastic and inelastic cross sections.

\subsubsection{Dependence of $V_{LS}$ \label{sec:LS}}

Here, we show the dependence of $V_{\rm LS}$ in the Hamiltonian of the nuclear structure calculation on the obtained nuclear structure and reaction.
The results are essentially the same as in Ref.~\cite{FUR21-2}.
However, we show the calculation results as a table instead of a figure in this paper.

\begin{table*}[ht]
\caption{Binding energies (BE), point-proton radius ($r_p$), point-neutron radius ($r_n$), point-nucleon-matter radius ($r_m$), expectation values of $\langle \bm{L} \cdot \bm{S} \rangle$ and $\langle \bm{S}^2 \rangle$ for the neutron, and transition strengths ($B$(E2), the neutron part ($B$(E2)$_n$), $B$(IS2), and $B$(IV2)) for $V_{\rm LS}$ = 0--4000 MeV.}
\label{tab:LS}
\begin{tabular}{lccccccccc} \hline \hline
$V_{\rm LS}$ (MeV) & 0 & 500 & 1000 & 1500 & 2000 & 2500 & 3000 &3500 & 4000 \\ \hline 
BE (0$^+_1$) (MeV) & -59.29 & -59.54 & -60.27 & -61.43 & -62.97 & -64.88 & -67.17 & -69.81 & -72.80  \\ 
BE (2$^+_1$) (MeV) & -57.14 & -57.23 & -57.57 & -58.37 & -59.78 & -61.66 & -63.93 & -66.57 & -69.52  \\
BE (2$^+_2$) (MeV) & -55.52 & -55.79 & -56.44 & -57.14 & -57.72 & -58.32 & -59.00 & -59.80 & -60.72  \\ \hline
$r_p$ (0$^+_1$) (fm) & 2.548 & 2.535 & 2.498 & 2.448 & 2.393 & 2.338 & 2.286 & 2.238 & 2.197  \\
$r_p$ (2$^+_1$) (fm) & 2.521 & 2.518 & 2.503 & 2.451 & 2.380 & 2.317 & 2.261 & 2.214 & 2.175  \\
$r_p$ (2$^+_2$) (fm) & 2.546 & 2.526 & 2.489 & 2.476 & 2.474 & 2.465 & 2.450 & 2.432 & 2.413  \\
$r_n$ (0$^+_1$) (fm) & 2.815 & 2.797 & 2.746 & 2.677 & 2.598 & 2.517 & 2.440 & 2.371 & 2.310  \\
$r_n$ (2$^+_1$) (fm) & 2.791 & 2.787 & 2.765 & 2.693 & 2.597 & 2.508 & 2.430 & 2.365 & 2.311  \\
$r_n$ (2$^+_2$) (fm) & 2.812 & 2.791 & 2.749 & 2.737 & 2.737 & 2.726 & 2.709 & 2.688 & 2.665  \\
$r_m$ (0$^+_1$) (fm) & 2.712 & 2.695 & 2.650 & 2.588 & 2.518 & 2.447 & 2.380 & 2.319 & 2.266  \\
$r_m$ (2$^+_1$) (fm) & 2.686 & 2.682 & 2.663 & 2.599 & 2.512 & 2.433 & 2.364 & 2.306 & 2.258  \\
$r_m$ (2$^+_2$) (fm) & 2.709 & 2.688 & 2.648 & 2.636 & 2.635 & 2.625 & 2.609 & 2.589 & 2.567  \\ \hline
$\langle \bm{L} \cdot \bm{S} \rangle$ (0$^+_1$) & -0.0002832 & 0.3766 & 0.6783 & 0.8871 & 1.025 & 1.118 & 1.182 & 1.226 & 1.257  \\
$\langle \bm{L} \cdot \bm{S} \rangle$ (2$^+_1$) & -0.0007262 & 0.09012 & 0.2824 & 0.6649 & 0.9164 & 1.038 & 1.114 & 1.167 & 1.203  \\
$\langle \bm{L} \cdot \bm{S} \rangle$ (2$^+_2$) & -0.002162 & 0.3583 & 0.4459 & 0.2394 & 0.1156 & 0.09424 & 0.09968 & 0.1135 & 0.1317  \\
$\langle \bm{S}^2 \rangle$ (0$^+_1$) & 0.001527 & 0.06294 & 0.2080 & 0.3717 & 0.5181 & 0.6376 & 0.7311 & 0.8028 & 0.8574  \\
$\langle \bm{S}^2 \rangle$ (2$^+_1$) & 0.002947 & 0.02264 & 0.1166 & 0.3520 & 0.5470 & 0.6632 & 0.7441 & 0.8049 & 0.8514  \\
$\langle \bm{S}^2 \rangle$ (2$^+_2$) & 0.002436 & 0.1035 & 0.2126 & 0.1503 & 0.09792 & 0.1056 & 0.1363 & 0.1785 & 0.2284  \\ \hline
$B$(E2: 2$^+_1$ $\to$ 0$^+_1$) ($e^2$ fm$^4$) & 2.952 & 3.615 & 6.473 & 11.14 & 11.82 & 10.61 & 9.222 & 8.005 & 7.006  \\
$B$((E2)$_n$: 2$^+_1$ $\to$ 0$^+_1$) (fm$^4$) & 18.12 & 18.47 & 19.47 & 17.46 & 12.72 & 9.508 & 7.411 & 5.944 & 4.883  \\  
$B$(IS2: 2$^+_1$ $\to$ 0$^+_1$) (fm$^4$) & 35.69 & 38.42 & 48.39 & 56.49 & 49.06 & 40.20 & 33.17 & 27.75 & 23.59 \\  
$B$(IV2: 2$^+_1$ $\to$ 0$^+_1$) (fm$^4$) & 6.442 & 5.740 & 3.491 & 0.7058 & 0.01631 & 0.03004 & 0.09894 & 0.1531 & 0.1911  \\  
$B$(E2: 2$^+_2$ $\to$ 0$^+_1$) ($e^2$ fm$^4$) & 16.14 & 14.56 & 9.867 & 3.228 & 0.6455 & 0.1297 & 0.01963 & 0.0001363 & 0.006089  \\
$B$((E2)$_n$: 2$^+_2$ $\to$ 0$^+_1$) (fm$^4$) & 5.861 & 4.443 & 1.108 & 0.4739 & 2.640 & 3.521 & 3.620 & 3.448 & 3.216  \\  
$B$(IS2: 2$^+_2$ $\to$ 0$^+_1$) (fm$^4$) & 41.46 & 35.09 & 17.59 & 1.228 & 0.6747 & 2.299 & 3.106 & 3.404 & 3.502  \\  
$B$(IV2: 2$^+_2$ $\to$ 0$^+_1$) (fm$^4$) & 2.549 & 2.917 & 4.363 & 6.175 & 5.897 & 5.002 & 4.173 & 3.491 & 2.943   \\  
$B$(E2: 2$^+_2$ $\to$ 2$^+_1$) ($e^2$ fm$^4$) & 12.75 & 15.30 & 22.45 & 15.92 & 4.390 & 1.114 & 0.2748 & 0.05108 & 0.001385  \\
$B$((E2)$_n$: 2$^+_2$ $\to$ 2$^+_1$) (fm$^4$) & 0.5122 & 1.971 & 12.25 & 25.59 & 17.21 & 10.17 & 6.362 & 4.169 & 2.807  \\
$B$(IS2: 2$^+_2$ $\to$ 2$^+_1$) (fm$^4$) & 18.37 & 28.25 & 67.87 & 81.88 & 38.98 & 18.01 & 9.281 & 5.143 & 2.933  \\
$B$(IV2: 2$^+_2$ $\to$ 2$^+_1$) (fm$^4$) & 8.149 & 6.288 & 1.531 & 1.142 & 4.213 & 4.551 & 3.992 & 3.297 & 2.684  \\\hline \hline
  \end{tabular}
\end{table*}

Table~\ref{tab:LS} shows the calculated binding energies (BE).
The experimental data are -65.07, -61.70, and -59.11 MeV for 0$^+_1$, 2$^+_1$, and 2$^+_2$, respectively.
By changing the strength $V_{\rm LS}$ in the range 0--4000 MeV, the binding energies are drastically changed.
A large $V_{\rm LS}$ value results in strong binding energy.
On the other hand, a small $V_{\rm LS}$ value results in weak binding energy.
The calculated values were almost identical to the experimental data around 2500 MeV.
However, we have taken 2000 MeV as a criterion for $V_{\rm LS}$, because the effect of the $\alpha + t + t$ cluster structure is not included in this paper.
The effect of the $\alpha$ cluster breaking and recombination gives a deeper binding energy~\cite{ITA08}.

Table~\ref{tab:LS} also shows the calculated root mean squared radii for the point proton ($r_p$), point neutron ($r_n$), and point nucleon matter ($r_m$) of the 0$^+_1$, 2$^+_1$, and 2$^+_2$ states.
The experimental values of the proton and matter radii of the $^{10}$Be ground state are 2.357 $\pm$ 0.018 fm~\cite{NOR09} and 2.39 $\pm$ 0.02 fm~\cite{TAN85}, respectively.
The calculated radii are very close to the experimental data around $V_{\rm LS}$ = 2500 MeV.
The increase in the strength of the spin-orbit interaction is responsible for the small size of the point nucleon-matter, neutron, and proton because of the strong binding.
The size of the 0$^+_1$ and 2$^+_1$  states show consistent behaviors, especially with the strong binding.
On the other hand, the 2$^+_2$ state, which is larger than the 0$^+_1$ and 2$^+_1$ states except for very small $V_{\rm LS}$ values, exhibits much gentler shrinkage behavior with increasing binding energy.
This result implies that the structure of the 2$^+_2$ state is quite different from that of the ground and 2$^+_1$ states when strong spin-orbit interaction is applied.
Indeed, the 2$^+_2$ state with $K=2$ corresponds to the excitation of one neutron from the spin-orbit favored orbit to the unfavored one according to the simple molecular-orbital picture.

In order to gain more precise information, we present the expectation values of $\langle \bm{L} \cdot \bm{S} \rangle$ and $\langle \bm{S}^2 \rangle$ for the neutron in Table~\ref{tab:LS}.
The expectation values of the proton part are 0 because all protons are in the $\alpha$ clusters.
The 0$^+_1$ and 2$^+_1$ states clearly exhibit similar growth of the expectation value with changing the strength of the spin-orbit interaction.
Thus, the $0^+_1$ and $2^+_1$ states have similar inner structures and undergo similar changes with increasing strength of the spin-orbit interaction.
The expectation value of the 2$^+_2$ state shows a different trend, unlike the behavior of the $0^+_1$ and $2^+_1$ states.
This finding implies that the inner structure of the 2$^+_2$ state is different from the structures of the 0$^+_1$ and 2$^+_1$ states when the strong spin-orbit interaction is applied.

Next, we see the quadrupole transition strengths between each state.
Table~\ref{tab:LS} also shows the quadrupole transition strengths ($B$(E2), the neutron part $B$((E2)$_n$), $B$(IS2), and $B$(IV2)) from the 2$^+_1$ state to the 0$^+_1$ state, from the 2$^+_2$ state to the 0$^+_1$ state, and from the 2$^+_2$ state to the 2$^+_1$ state.
Although the radii and expectation values of the 0$^+_1$ and 2$^+_1$ states show similar tendencies, the change in the transition strengths of $B$(E2) and $B$((E2)$_n$) from the 2$^+_1$ state to the 0$^+_1$ state is not simple.
Under weak spin-orbit interaction, the strengths of the proton and neutron parts are quite different.
With increasing the $V_{\rm LS}$ value, the transition strength of the proton part becomes large.
For $V_{\rm LS}$ = 2000--4000 MeV, both the proton and neutron parts decrease together.
This behavior is caused by the sign reversal of the transition density for the neutron part as mentioned in Ref.~\cite{FUR21-2}.
From another perspective, we see IS and IV transition strengths.
The obtained IS transition strengths from 2$^+_1$ to 0$^+_1$ stay in the range of 20--60~fm$^4$.
The IV transition strengths from 2$^+_1$ to 0$^+_1$ get gradually small as $V_{\rm LS}$ grows.
On the other hand, the IS transition strengths from 2$^+_2$ to 0$^+_1$ are drastically changed by the $V_{\rm LS}$ value.
This effect on the cross section will be discussed later.
Here, we also discuss the transition strength from the 2$^+_2$ state to the 2$^+_1$ state.
The $B$(E2: 2$^+_2$ $\to$ 2$^+_1$) value becomes smaller when the strength of the $V_{\rm LS}$ value becomes larger.
In Ref.~\cite{ITA02}, it is mentioned that the increase of the $B$(E2: 2$^+_2$ $\to$ 2$^+_1$) value with the decrease of the $V_{\rm LS}$ value supports a triaxial $\alpha$ + $\alpha$ + dineutron clustering configuration.
However, the $B$(E2: 2$^+_2$ $\to$ 2$^+_1$) values with $V_{\rm LS}$ = 0 and 500 MeV are smaller than that with 1000 MeV.
This is considered to be caused by the further development of the dineutron cluster.
Namely, the coupling pattern of the angular momentum among the two $\alpha$ clusters and the dineutron cluster is different in the 2$^+_1$ and 2$^+_2$ states.
Concretely, when the small $V_{\rm LS}$ value is adopted, it is considered that the two $\alpha$ clusters with the relative angular momentum $L=0$ are synthesized with the dineutron around them with $L=2$ in the 2$^+_1$ state. 
On the other hand, the two $\alpha$ clusters with the relative angular momentum $L=2$ are synthesized with the dineutron moving around them with $L=0$ in the 2$^+_2$ state.
The decrease of the $B$(E2: 2$^+_2$ $\to$ 2$^+_1$) value and the increase of the $B$(IV2: 2$^+_2$ $\to$ 2$^+_1$) value corroborate the situation.
We will discuss this situation again in Sec.~\ref{sec:dineutron}.

\begin{figure}[ht]
\centering
\includegraphics[width=6.4cm]{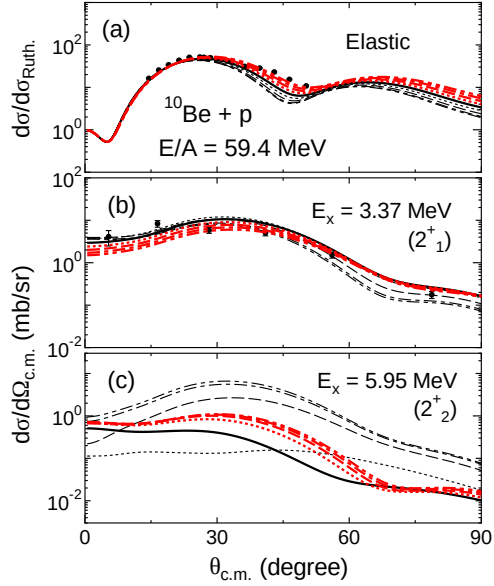}
\caption{\label{fig:LS-59.4mev}
(a) Elastic cross section, (b) inelastic cross section for the 2$^+_1$ state, and (c) inelastic cross section for the 2$^+_2$ state of $^{10}$Be + p system at $E/A =$ 59.4 MeV.
The detail for the curves is written in the text.
The experimental data are taken from \cite{EXFOR, COR97, IWA00}.}
\end{figure}

\begin{figure}[ht]
\centering
\includegraphics[width=6.4cm]{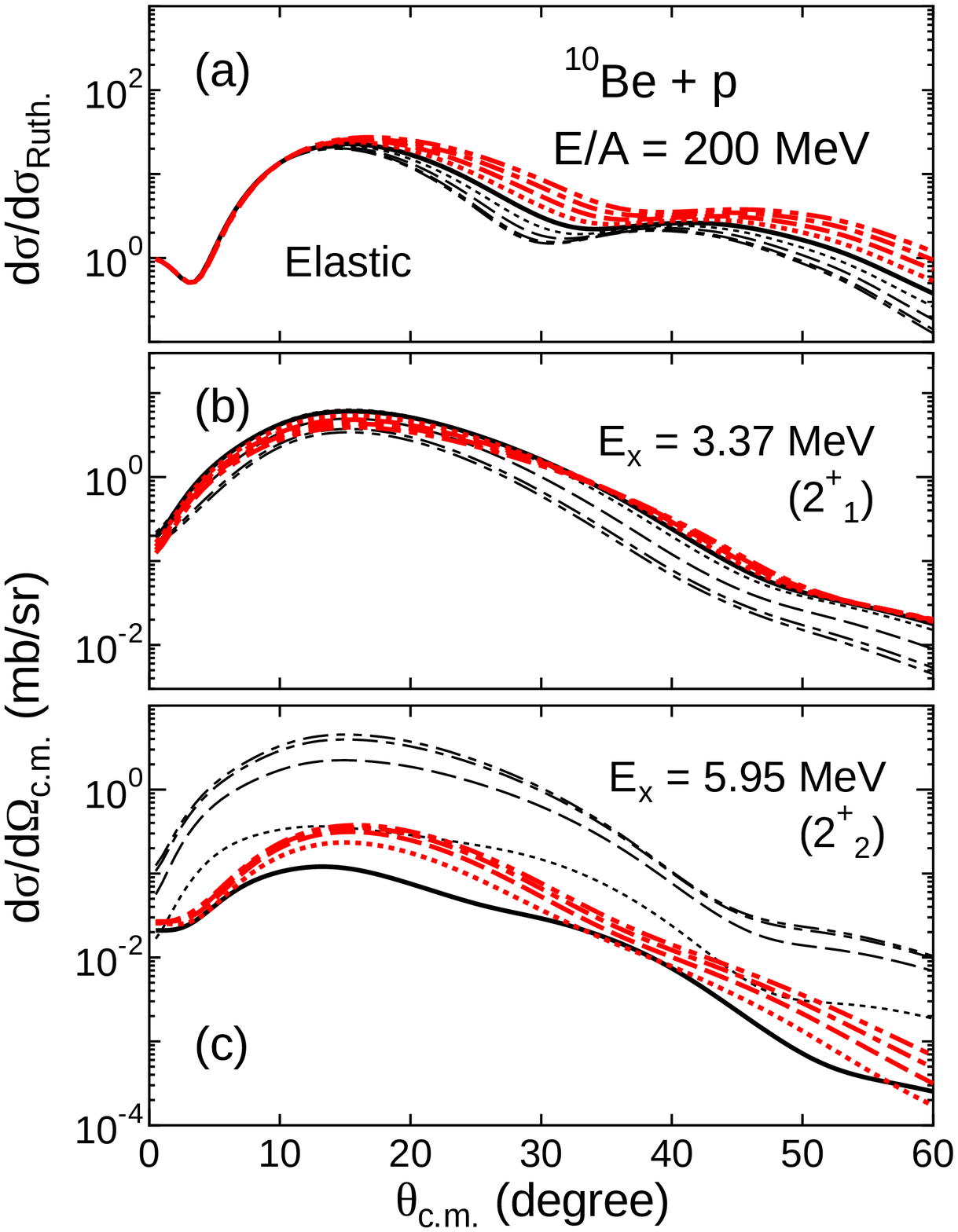}
\caption{\label{fig:LS-200mev}
Same as Fig.~\ref{fig:LS-59.4mev} but at $E/A =$ 200 MeV.}
\end{figure}

\begin{figure}[ht]
\centering
\includegraphics[width=6.4cm]{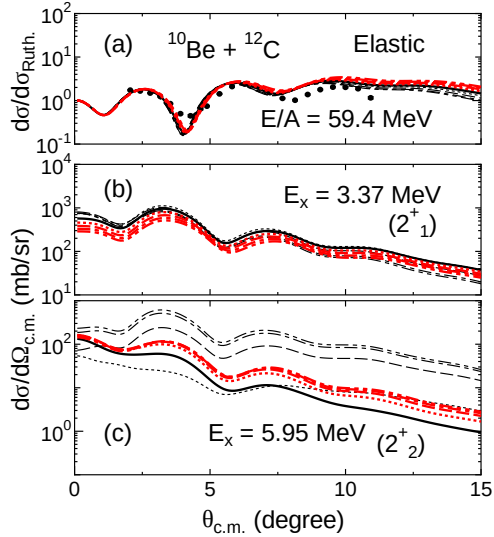}
\caption{\label{fig:LS2-59.4mev}
Same as Fig.~\ref{fig:LS-59.4mev} but for the $^{12}$C target.
The experimental data are taken from \cite{EXFOR,COR97}.}
\end{figure}

\begin{figure}[ht]
\centering
\includegraphics[width=6.4cm]{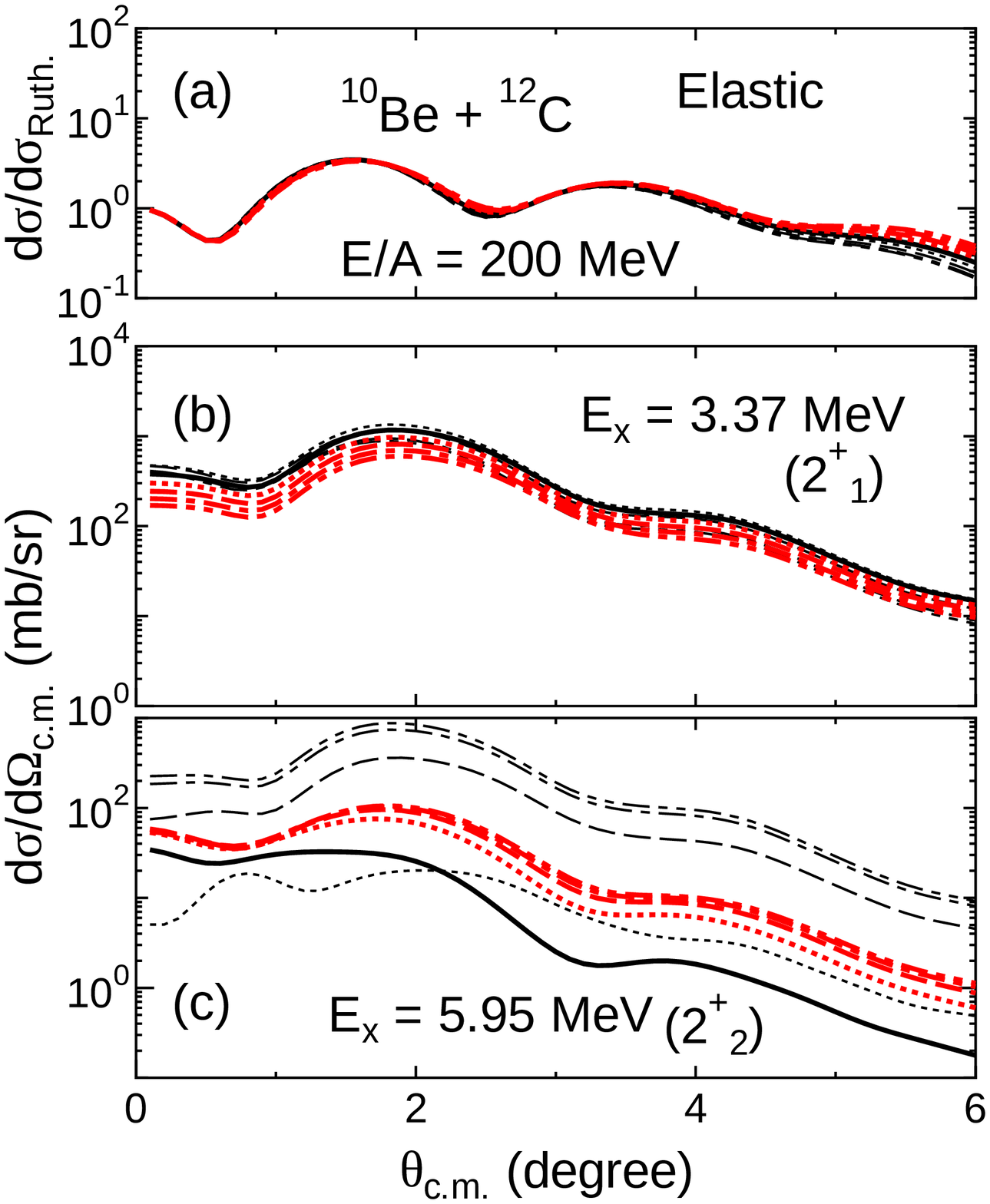}
\caption{\label{fig:LS2-200mev}
Same as Fig.~\ref{fig:LS2-59.4mev} but at $E/A =$ 200 MeV.}
\end{figure}

Figure~\ref{fig:LS-59.4mev} shows the (a) elastic cross section, (b) inelastic cross section for the 2$^+_1$ state, and (c) inelastic cross section for the 2$^+_2$ state of the $^{10}$Be nucleus by the proton target at $E/A =$ 59.4 MeV.
The two-dot-dashed, dot-dashed, dashed, dotted, bold-solid, bold(red)-dotted, bold(red)-dashed, bold(red)-dot-dashed, and bold(red)-two-dot-dashed curves are obtained with $V_{LS}$ = 0, 500, 1000, 1500, 2000, 2500, 3000, 3500, and 4000 MeV, respectively.
The experimental data are well reproduced in a wide range of the $V_{\rm LS}$ values.
The change in the size of the ground state with varying $V_{\rm LS}$ slightly affects the elastic cross section.
The relation among the binding energy, the nuclear size, and the elastic cross section will be discussed in Sec.~\ref{sec:BE} in detail.
The visible change in the transition strength from the 2$^+_1$ state to the 0$^+_1$ state for the proton and neutron parts results in a slight change in the inelastic cross section.
As mentioned above, the $B$(IS2) values are obtained within 20--60 fm$^4$.
Therefore, the calculated inelastic cross sections are similar to each other.
On the other hand, the calculated inelastic cross sections for the 2$^+_2$ state are greatly dependent on the $V_{\rm LS}$ value adopted in the structure calculation.
The drastic change in the inelastic cross section is caused by the difference in the transition strengths, as shown in Table~\ref{tab:LS}.
As mentioned in Ref.~\cite{FUR21-2}, the behavior of the neutron part, in particular, is important and affects the IS transition strength from the 2$^+_2$ state to the 0$^+_1$ state.
Here, we mention that the transition strength from the 2$^+_2$ state to the 2$^+_1$ state cannot be ignored.
Therefore, the multistep reaction effect (0$^+_1$ $\to$ 2$^+_1$ $\to$ 2$^+_2$) also contributes and makes the inelastic cross section for the 2$^+_2$ state at $E/A =$ 59.4 MeV complicated. 
To avoid the multistep reaction effect, we also calculate the elastic and inelastic cross sections at a higher energy at which the multistep reaction effect is considered to be minor.
In fact, we confirmed that the multistep reaction effect arising from the transition from the 2$^+_1$ state to the 2$^+_2$ state is minor at 200~MeV.
Figure~\ref{fig:LS-200mev} shows the calculated (a) elastic cross section, (b) inelastic cross section for the 2$^+_1$ state, and (c) inelastic cross section for the 2$^+_2$ state at $E/A =$ 200~MeV.
The inelastic cross section for the 2$^+_2$ state is drastically influenced by the choice of the spin-orbit strength for the structure calculation, and this choice affects the inner structure of $^{10}$Be.
On the other hand, it is difficult to distinguish the effect of the $V_{\rm LS}$ parameter on the elastic cross section and the inelastic cross section for the 2$^+_1$ state.

We also calculate the elastic and inelastic cross sections for the $^{10}$Be + $^{12}$C system at $E/A=$ 59.4 and 200 MeV to investigate the effect of a different target.
Figures~\ref{fig:LS2-59.4mev} and \ref{fig:LS2-200mev} show the (a) elastic cross section, (b) inelastic cross section for the 2$^+_1$ state, and (c) inelastic cross section for the 2$^+_2$ state of the $^{10}$Be + $^{12}$C system at $E/A =$ 59.4 and 200 MeV.
The difference in the nuclear size has a minor effect on the elastic cross sections for the $^{12}$C target.
The inelastic cross sections for the 2$^+_1$ state for the $^{12}$C target have similar behavior to each other.
Also, we can see the drastic change of the inelastic cross sections for the 2$^+_2$ state.
Here, we note that the calculated inelastic cross sections for the 2$^+_2$ state show different behavior in comparison with the proton target, especially for the dotted and bold solid curves.
The results are caused by the weak IS transition strength shown in Table~\ref{tab:LS}.
In the present calculation for the $^{12}$C target, the IV component has no effect on the cross section.
Comparing the proton target with $^{12}$C, the effect of IS and IV components can be seen.

\subsubsection{Dependence of $M$ \label{sec:M}}

Next, we investigate the dependence of the Majorana $(M)$ parameter.
The $M$ parameter is responsible for describing the nucleon-nucleon exchange effect and is related to the Wigner ($W$) parameter as $W=1-M$.
The values of 0.56--0.63 are widely used in the cluster model calculations.
For the $^{10}$Be nucleus, 0.56 or 0.60 is usually adopted.
Therefore, we investigate the range of $M$ = 0.52--0.64, which is wider than the usual cases, in this paper.

\begin{table*}[ht]
\caption{Same as Table~\ref{tab:LS} but for $M$ = 0.52--0.64.}
\label{tab:M}
\begin{tabular}{lccccccc} \hline \hline
$M$ & 0.52 & 0.54 & 0.56 & 0.58 & 0.60 & 0.62 & 0.64 \\ \hline 
BE (0$^+_1$) (MeV) & -77.07 & -73.08 & -69.39 & -66.01 & -62.97 & -60.26 & -57.90  \\
BE (2$^+_1$) (MeV) & -74.22 & -70.17 & -66.38 & -62.91 & -59.78 & -57.05 & -54.73  \\
BE (2$^+_2$) (MeV) & -69.52 & -66.18 & -63.11 & -60.29 & -57.72 & -55.40 & -53.34  \\ \hline
$r_p$ (0$^+_1$) (fm) & 2.160 & 2.205 & 2.258 & 2.320 & 2.393 & 2.479 & 2.579  \\
$r_p$ (2$^+_1$) (fm) & 2.139 & 2.180 & 2.231 & 2.296 & 2.380 & 2.488 & 2.617  \\
$r_p$ (2$^+_2$) (fm) & 2.237 & 2.286 & 2.341 & 2.404 & 2.474 & 2.557 & 2.660  \\
$r_n$ (0$^+_1$) (fm) & 2.298 & 2.357 & 2.426 & 2.506 & 2.598 & 2.702 & 2.816  \\
$r_n$ (2$^+_1$) (fm) & 2.292 & 2.346 & 2.411 & 2.494 & 2.597 & 2.722 & 2.863  \\
$r_n$ (2$^+_2$) (fm) & 2.446 & 2.509 & 2.579 & 2.655 & 2.737 & 2.828 & 2.933  \\
$r_m$ (0$^+_1$) (fm) & 2.244 & 2.297 & 2.360 & 2.433 & 2.518 & 2.615 & 2.724  \\
$r_m$ (2$^+_1$) (fm) & 2.232 & 2.281 & 2.341 & 2.416 & 2.512 & 2.631 & 2.767  \\
$r_m$ (2$^+_2$) (fm) & 2.365 & 2.423 & 2.487 & 2.558 & 2.635 & 2.723 & 2.827  \\ \hline
$\langle \bm{L} \cdot \bm{S} \rangle$ (0$^+_1$) & 1.210 & 1.177 & 1.136 & 1.085 & 1.025 & 0.9571 & 0.8816  \\
$\langle \bm{L} \cdot \bm{S} \rangle$ (2$^+_1$) & 1.121 & 1.091 & 1.050 & 0.9938 & 0.9164 & 0.8123 & 0.6919  \\
$\langle \bm{L} \cdot \bm{S} \rangle$ (2$^+_2$) & 0.08979 & 0.08770 & 0.08810 & 0.1038 & 0.1156 & 0.1560 & 0.2073  \\
$\langle \bm{S}^2 \rangle$ (0$^+_1$) & 0.7115 & 0.6723 & 0.6271 & 0.5755 & 0.5181 & 0.4560 & 0.3915  \\
$\langle \bm{S}^2 \rangle$ (2$^+_1$) & 0.7017 & 0.6742 & 0.6416 & 0.6011 & 0.5470 & 0.4731 & 0.3835  \\
$\langle \bm{S}^2 \rangle$ (2$^+_2$) & 0.1009 & 0.09268 & 0.08663 & 0.08654 & 0.09792 & 0.1241 & 0.1542  \\ \hline
$B$(E2: 2$^+_1$ $\to$ 0$^+_1$) ($e^2$ fm$^4$) & 6.049 & 7.000 & 8.226 & 9.811 & 11.82 & 14.26 & 17.17  \\
$B$((E2)$_n$: 2$^+_1$ $\to$ 0$^+_1$) (fm$^4$) & 3.916 & 4.981 & 6.557 & 8.970 & 12.72 & 18.35 & 26.07  \\  
$B$(IS2: 2$^+_1$ $\to$ 0$^+_1$) (fm$^4$) & 19.70 & 23.79 & 29.47 & 37.54 & 49.06 & 64.96 & 85.55 \\  
$B$(IV2: 2$^+_1$ $\to$ 0$^+_1$) (fm$^4$) & 0.2309 & 0.1713 & 0.09447 & 0.01885 & 0.01631 & 0.2579 & 0.9271  \\  
$B$(E2: 2$^+_2$ $\to$ 0$^+_1$) ($e^2$ fm$^4$) & 0.02127 & 0.04007 & 0.09132 & 0.2363 & 0.6455 & 1.704 & 3.907  \\
$B$((E2)$_n$: 2$^+_2$ $\to$ 0$^+_1$) (fm$^4$) & 2.793 & 3.039 & 3.202 & 3.141 & 2.640 & 1.592 & 0.4369  \\  
$B$(IS2: 2$^+_2$ $\to$ 0$^+_1$) (fm$^4$) & 2.327 & 2.382 & 2.212 & 1.654 & 0.6747 & 0.001903 & 1.731  \\  
$B$(IV2: 2$^+_2$ $\to$ 0$^+_1$) (fm$^4$) & 3.302 & 3.777 & 4.375 & 5.100 & 5.897 & 6.589 & 6.958   \\  
$B$(E2: 2$^+_2$ $\to$ 2$^+_1$) ($e^2$ fm$^4$) & 0.09201 & 0.2346 & 0.6204 & 1.666 & 4.390 & 10.60 & 21.48  \\
$B$((E2)$_n$: 2$^+_2$ $\to$ 2$^+_1$) (fm$^4$) & 4.574 & 5.816 & 7.856 & 11.32 & 17.21 & 26.41 & 38.68  \\
$B$(IS2: 2$^+_2$ $\to$ 2$^+_1$) (fm$^4$) & 5.963 & 8.387 & 12.89 & 21.68 & 38.98 & 70.46 & 117.8  \\
$B$(IV2: 2$^+_2$ $\to$ 2$^+_1$) (fm$^4$) & 3.368 & 3.715 & 4.061 & 4.302 & 4.213 & 3.548 & 2.513  \\\hline \hline
  \end{tabular}
\end{table*}

Table~\ref{tab:M} shows calculated values of binding energies, radii, expectation values, and quadrupole transition strengths.
Not only the $V_{\rm LS}$ parameter but also the $M$ parameter give  drastic changes in the binding energies.
With the increase of $M$, the repulsive effect becomes larger.
Therefore, the large (small) $M$ value gives weak (strong) binding.
However, it is found that $M$ has a minor effect on the excitation energy of the 2$^+_1$ state.
On the contrary, when the $M$ parameter is large, the difference of the energies between the 2$^+_1$ and 2$^+_2$ states becomes small.
The size of the ground and excited states are also influenced by the change in the binding energy.
The strong binding gives the small size for the ground and excited states.
Therefore, the trend of the dependence of size on the $M$ parameter is simple.
Also, the expectation values show a simple increase or decrease, except for the $\langle \bm{S}^2 \rangle$ value of 2$^+_2$ state, which shows minor dependence on the $M$ value.

Next, we discuss the transition strength.
Both of the proton and neutron parts of the transition strengths from the 2$^+_1$ state to the 0$^+_1$ state simply increase with the $M$ parameter.
The increase is concerned with the IS component.
For the transition strength from the 2$^+_2$ state to the 0$^+_1$ state, the proton part increases, but the neutron part decreases with the $M$ parameter.
The increase of the proton part and the decrease of the neutron part give complicated transition strength of the IS and IV components.
The IV component of the transition strength from the 2$^+_2$ state to the 0$^+_1$ state is always larger than the IS component.
This result is also different in comparison with the $V_{\rm LS}$ case.
In addition, the transition strength from the 2$^+_2$ state to the 2$^+_1$ state gets larger with the $M$ parameter.
This indicates that we cannot ignore the multistep reaction effect on the inelastic cross section caused by the strong transition from the 2$^+_1$ state to the 2$^+_2$ state.

\begin{figure}[ht]
\centering
\includegraphics[width=6.4cm]{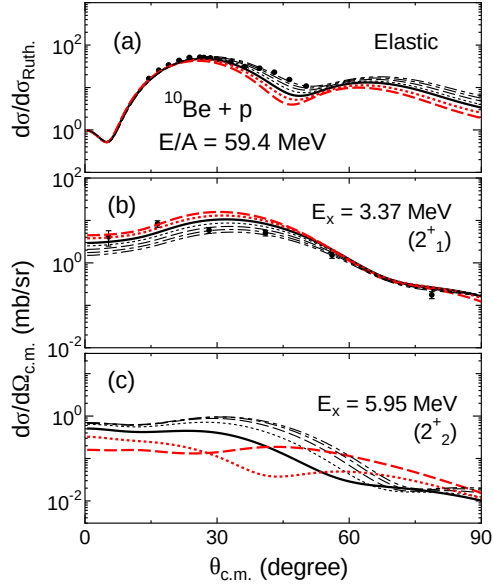}
\caption{\label{fig:M-59.4mev}
Same as Fig.~\ref{fig:LS-59.4mev} but for $M$ = 0.52--0.64.
The detail is written in the text.}
\end{figure}

\begin{figure}[ht]
\centering
\includegraphics[width=6.4cm]{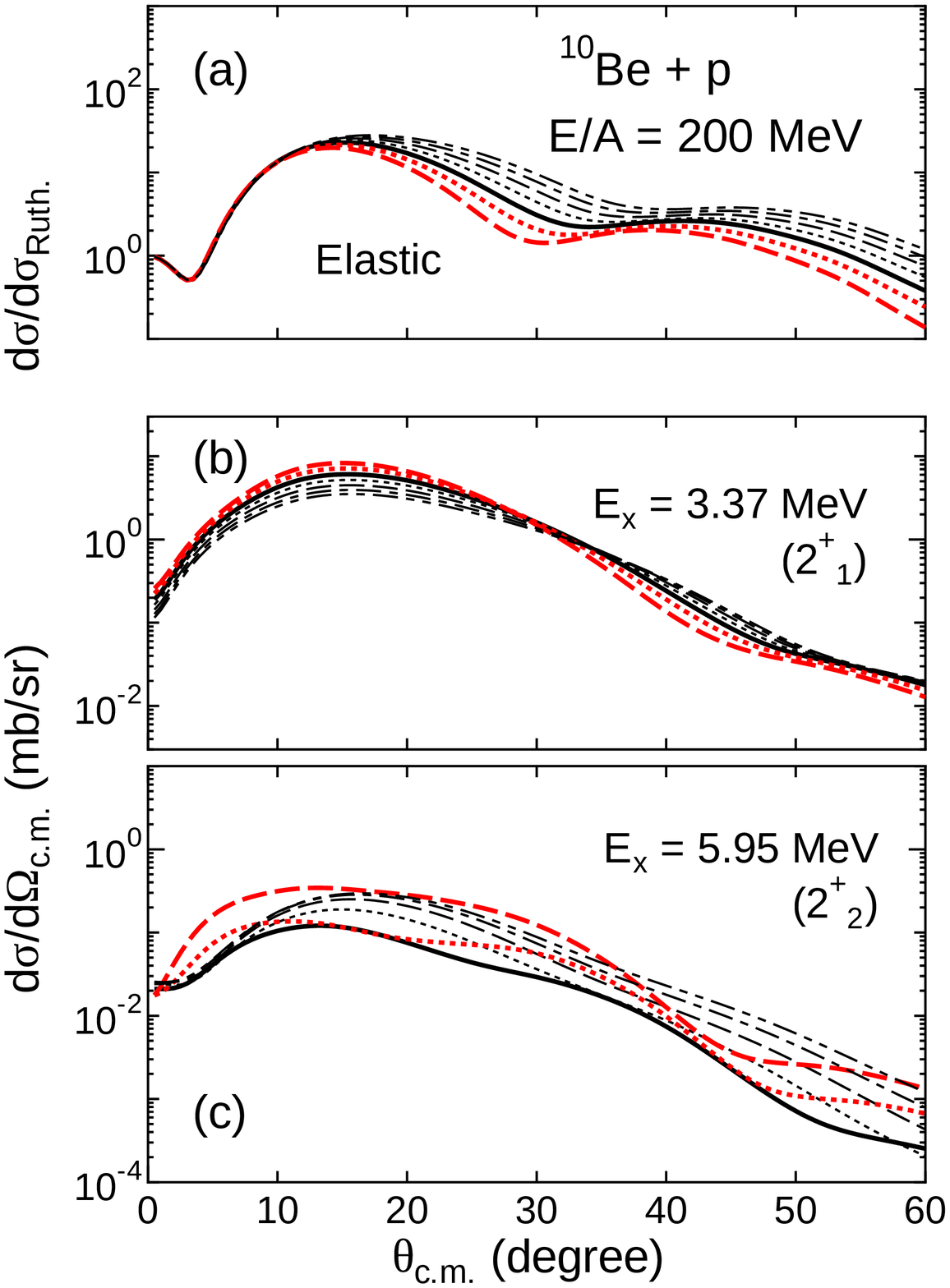}
\caption{\label{fig:M-200mev}
Same as Fig.~\ref{fig:M-59.4mev} but at $E/A$ = 200 MeV.}
\end{figure}

\begin{figure}[ht]
\centering
\includegraphics[width=6.4cm]{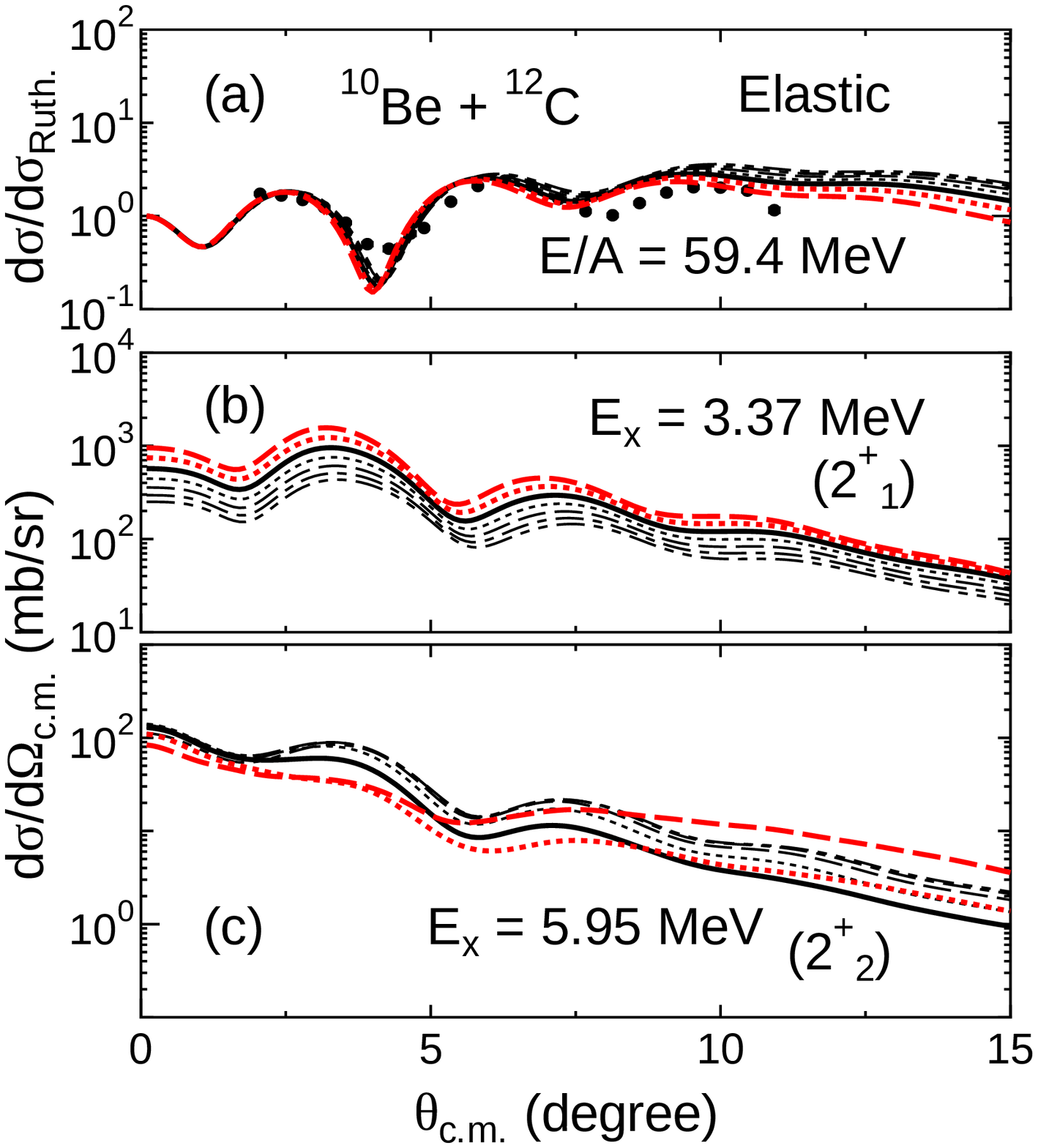}
\caption{\label{fig:M2-59.4mev}
Same as Fig.~\ref{fig:M-59.4mev} but for the $^{12}$C target.}
\end{figure}

\begin{figure}[ht]
\centering
\includegraphics[width=6.4cm]{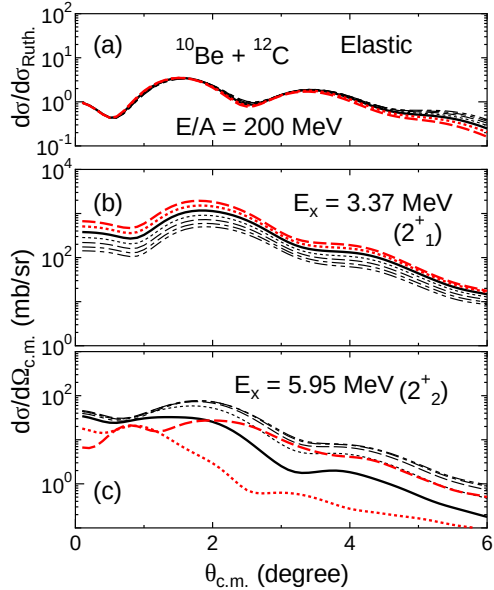}
\caption{\label{fig:M2-200mev}
Same as Fig.~\ref{fig:M-200mev} but for the $^{12}$C target.}
\end{figure}

Figures~\ref{fig:M-59.4mev} and \ref{fig:M-200mev} show the calculated elastic and inelastic cross sections by the proton target with the experimental data.
The two-dot-dashed, dot-dashed, dashed, dotted, bold-solid, bold(red)-dotted, and bold(red)-dashed curves are obtained with $M$ = 0.52, 0.54, 0.56, 0.58, 0.60, 0.62, and 0.64, respectively.
The effect of changing the nuclear size is barely visible on the elastic cross section.
We will discuss the effect of the binding energy, nuclear size, and elastic cross section again in Sec.~\ref{sec:BE}.
The effect of the change of the transition strength from the 2$^+_1$ state to the 0$^+_1$ state is barely visible on the inelastic cross section for the 2$^+_1$ state.
On the contrary, we can see the drastic change of the inelastic cross sections for the 2$^+_2$ state.
In Ref.~\cite{FUR21-2}, we expected that the change of the $M$ parameter would not have much impact on the inelastic cross section.
However, it indeed does give the change.
Here, we compare the inelastic cross sections for the 2$^+_2$ state at $E/A=$ 59.4 with 200 MeV.
At 59.4 MeV, the multistep reaction effect caused by the strong transition strength from the 2$^+_2$ state to the 2$^+_1$ state
gives complicated angular distribution.
On the other hand, we can discuss the effect of changing the $M$ parameter on the inelastic cross section for the 2$^+_2$ state at 200 MeV because the multistep reaction effect can be ignored at higher incident energy.
The effect of changing $M$ is indeed seen on the inelastic cross section for the 2$^+_2$ state at 200 MeV as shown in Fig.~\ref{fig:M-200mev}.
Although this is smaller compared with the change of the $V_{\rm LS}$ parameter shown in Fig.~\ref{fig:LS-200mev}, we cannot ignore the effect.
Therefore, we will discuss the contribution of the $V_{\rm LS}$ and $M$ parameters simultaneously in Sec.~\ref{sec:BE}.

Figures~\ref{fig:M2-59.4mev} and \ref{fig:M2-200mev} show the calculated elastic and inelastic cross sections by the $^{12}$C target with the experimental data.
Again, the change in the nuclear size has a minor effect on the elastic cross sections for the $^{12}$C target.
The inelastic cross sections for the 2$^+_1$ state for the $^{12}$C target show similar behavior to that for the proton target.
Also, we can see the drastic change of the inelastic cross sections for the 2$^+_2$ state.
However, the behavior of the inelastic cross sections obtained with weak IS transition to the 2$^+_2$ state is slightly different in comparison with the proton target.
This is caused by the small transition strength of the IS component.
Again, we note that the IV component has no effect on the present calculation for the $^{12}$C target.

\subsubsection{Dependence of $B, H$}

Next, we investigate the dependence of the $B$ and $H$ parameters.
The role of these parameters is to adjust the relative strength of the proton-proton (neutron-neutron) and proton-neutron interactions.
The original Volkov interaction has no $B$ and $H$ terms.
However, not only deuteron but also two neutrons are bound within $B=H=0$.
Therefore, the $B$ and $H$ are often applied to give a realistic condition with the Volkov interaction.

\begin{table*}[ht]
\caption{Same as Table~\ref{tab:LS} but for $B = H$ = 0--0.20.}
\label{tab:pbar}
\begin{tabular}{lcccccc} \hline \hline
$B = H $ & 0 & 0.04 & 0.08 & 0.12 & 0.16 & 0.20  \\ \hline 
BE (0$^+_1$) (MeV) & -63.85 & -63.39 & -62.97 & -62.57 & -62.20 & -61.85  \\
BE (2$^+_1$) (MeV) & -60.58 & -60.17 & -59.78 & -59.42 & -59.09 & -58.78  \\
BE (2$^+_2$) (MeV) & -58.49 & -58.10 & -57.72 & -57.37 & -57.03 & -56.71  \\ \hline
$r_p$ (0$^+_1$) (fm) & 2.398 & 2.396 & 2.393 & 2.391 & 2.390 & 2.388  \\
$r_p$ (2$^+_1$) (fm) & 2.387 & 2.383 & 2.380 & 2.378 & 2.377 & 2.376  \\
$r_p$ (2$^+_2$) (fm) & 2.470 & 2.472 & 2.474 & 2.477 & 2.480 & 2.482  \\
$r_n$ (0$^+_1$) (fm) & 2.591 & 2.594 & 2.598 & 2.602 & 2.606 & 2.611  \\
$r_n$ (2$^+_1$) (fm) & 2.591 & 2.593 & 2.597 & 2.601 & 2.605 & 2.611  \\
$r_n$ (2$^+_2$) (fm) & 2.715 & 2.726 & 2.737 & 2.749 & 2.761 & 2.773  \\
$r_m$ (0$^+_1$) (fm) & 2.516 & 2.517 & 2.518 & 2.520 & 2.522 & 2.525  \\
$r_m$ (2$^+_1$) (fm) & 2.511 & 2.511 & 2.512 & 2.514 & 2.516 & 2.520  \\
$r_m$ (2$^+_2$) (fm) & 2.620 & 2.627 & 2.635 & 2.643 & 2.652 & 2.660  \\ \hline
$\langle \bm{L} \cdot \bm{S} \rangle$ (0$^+_1$) & 0.9680 & 0.9975 & 1.025 & 1.052 & 1.077 & 1.099  \\
$\langle \bm{L} \cdot \bm{S} \rangle$ (2$^+_1$) & 0.8667 & 0.8924 & 0.9164 & 0.9381 & 0.9574 & 0.9739  \\
$\langle \bm{L} \cdot \bm{S} \rangle$ (2$^+_2$) & 0.1217 & 0.1189 & 0.1156 & 0.1121 & 0.1083 & 0.1045  \\
$\langle \bm{S}^2 \rangle$ (0$^+_1$) & 0.4484 & 0.4827 & 0.5181 & 0.5544 & 0.5913 & 0.6285  \\
$\langle \bm{S}^2 \rangle$ (2$^+_1$) & 0.4658 & 0.5055 & 0.5470 & 0.5900 & 0.6414 & 0.6791  \\
$\langle \bm{S}^2 \rangle$ (2$^+_2$) & 0.09248 & 0.09534 & 0.09792 & 0.1002 & 0.1022 & 0.1040  \\ \hline
$B$(E2: 2$^+_1$ $\to$ 0$^+_1$) ($e^2$ fm$^4$) & 11.98 & 11.90 & 11.82 & 11.76 & 11.71 & 11.67  \\
$B$((E2)$_n$: 2$^+_1$ $\to$ 0$^+_1$) (fm$^4$) & 13.11 & 12.90 & 12.72 & 12.54 & 12.39 & 12.25  \\  
$B$(IS2: 2$^+_1$ $\to$ 0$^+_1$) (fm$^4$) & 50.15 & 49.58 & 49.06 & 48.60 & 48.19 & 47.83 \\
$B$(IV2: 2$^+_1$ $\to$ 0$^+_1$) (fm$^4$) & 0.02527 & 0.02055 & 0.01631 & 0.01262 & 0.009488 & 0.006893 \\  
$B$(E2: 2$^+_2$ $\to$ 0$^+_1$) ($e^2$ fm$^4$) & 0.7458 & 0.6949 & 0.6455 & 0.5982 & 0.5534 & 0.5112  \\
$B$((E2)$_n$: 2$^+_2$ $\to$ 0$^+_1$) (fm$^4$) & 2.552 & 2.597 & 2.640 & 2.681 & 2.719 & 2.754  \\  
$B$(IS2: 2$^+_2$ $\to$ 0$^+_1$) (fm$^4$) & 0.5384 & 0.6051 & 0.6747 & 0.7464 & 0.8191 & 0.8919  \\
$B$(IV2: 2$^+_2$ $\to$ 0$^+_1$) (fm$^4$) & 6.056 & 5.978 & 5.897 & 5.812 & 5.726 & 5.638  \\  
$B$(E2: 2$^+_2$ $\to$ 2$^+_1$) ($e^2$ fm$^4$) & 4.965 & 4.677 & 4.390 & 4.108 & 3.834 & 3.569  \\
$B$((E2)$_n$: 2$^+_2$ $\to$ 2$^+_1$) (fm$^4$) & 18.18 & 17.70 & 17.21 & 16.70 & 16.18 & 15.66  \\
$B$(IS2: 2$^+_2$ $\to$ 2$^+_1$) (fm$^4$) & 42.15 & 40.57 & 38.98 & 37.37 & 35.77 & 34.17  \\
$B$(IV2: 2$^+_2$ $\to$ 2$^+_1$) (fm$^4$) & 4.144 & 4.180 & 4.213 & 4.241 & 4.262 & 4.275  \\\hline \hline
  \end{tabular}
\end{table*}

Table~\ref{tab:pbar} shows calculated values of binding energies, radii, expectation values, and quadrupole transition strengths.
Almost all the calculated values show a minor dependence on the change of the $B$ and $H$ parameters.
The expectation value of $\langle \bm{S}^2 \rangle$ for the 0$^+_1$, 2$^+_1$ and 2$^+_1$ states and the transition strength from the 2$^+_2$ state to the 0$^+_1$ state are slightly affected.
However, their causal relationship is not clear.

\begin{figure}[ht]
\centering
\includegraphics[width=6.4cm]{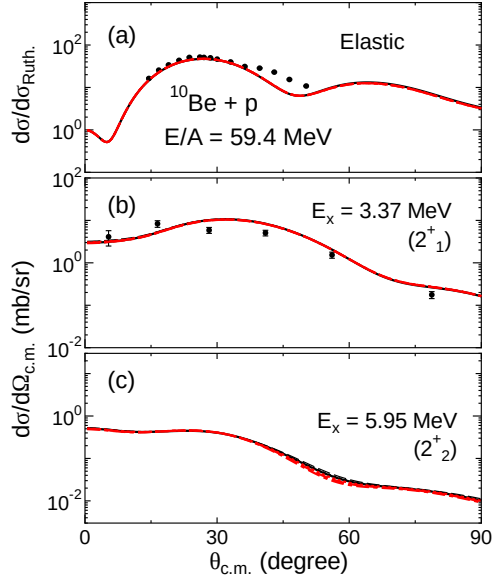}
\caption{\label{fig:pbar-59.4mev}
Same as Fig.~\ref{fig:LS-59.4mev} but with $B = H$ = 0--0.20.
The detail is written in the text.}
\end{figure}

\begin{figure}[ht]
\centering
\includegraphics[width=6.4cm]{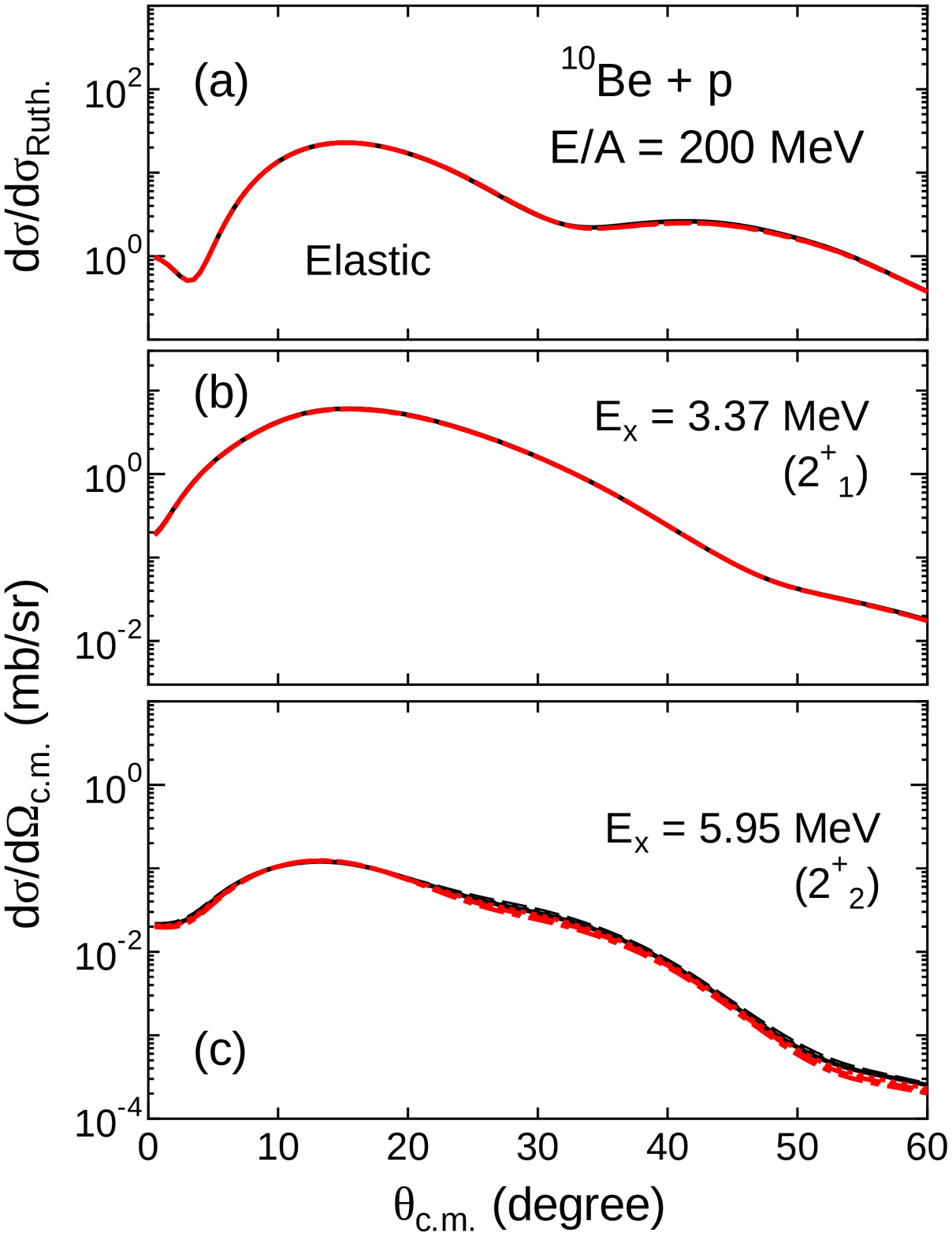}
\caption{\label{fig:pbar-200mev}
Same as Fig.~\ref{fig:pbar-59.4mev} but at $E/A$ = 200 MeV.}
\end{figure}

\begin{figure}[ht]
\centering
\includegraphics[width=6.4cm]{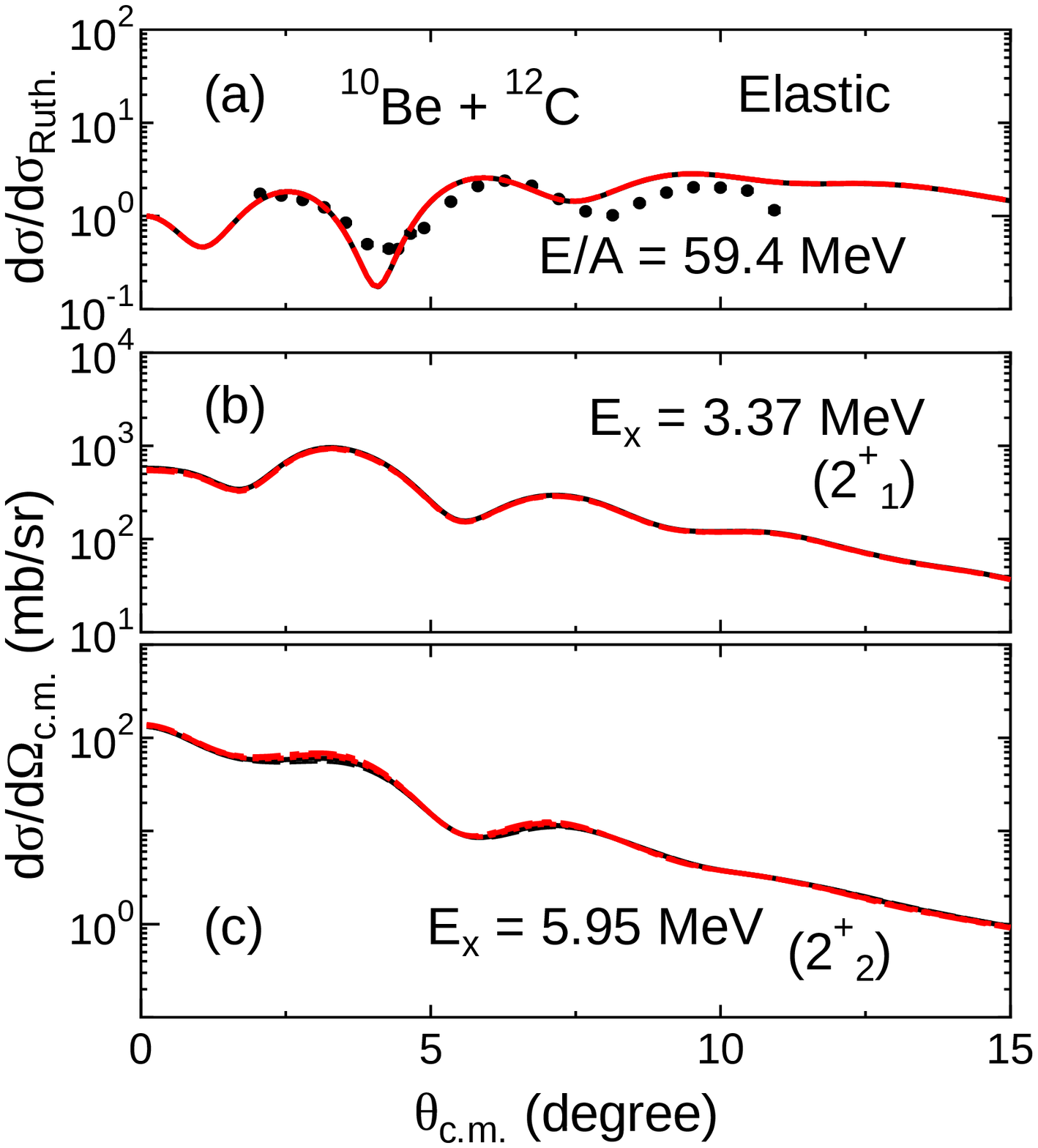}
\caption{\label{fig:pbar2-59.4mev}
Same as Fig.~\ref{fig:pbar-59.4mev} but for the $^{12}$C target.}
\end{figure}

\begin{figure}[ht]
\centering
\includegraphics[width=6.4cm]{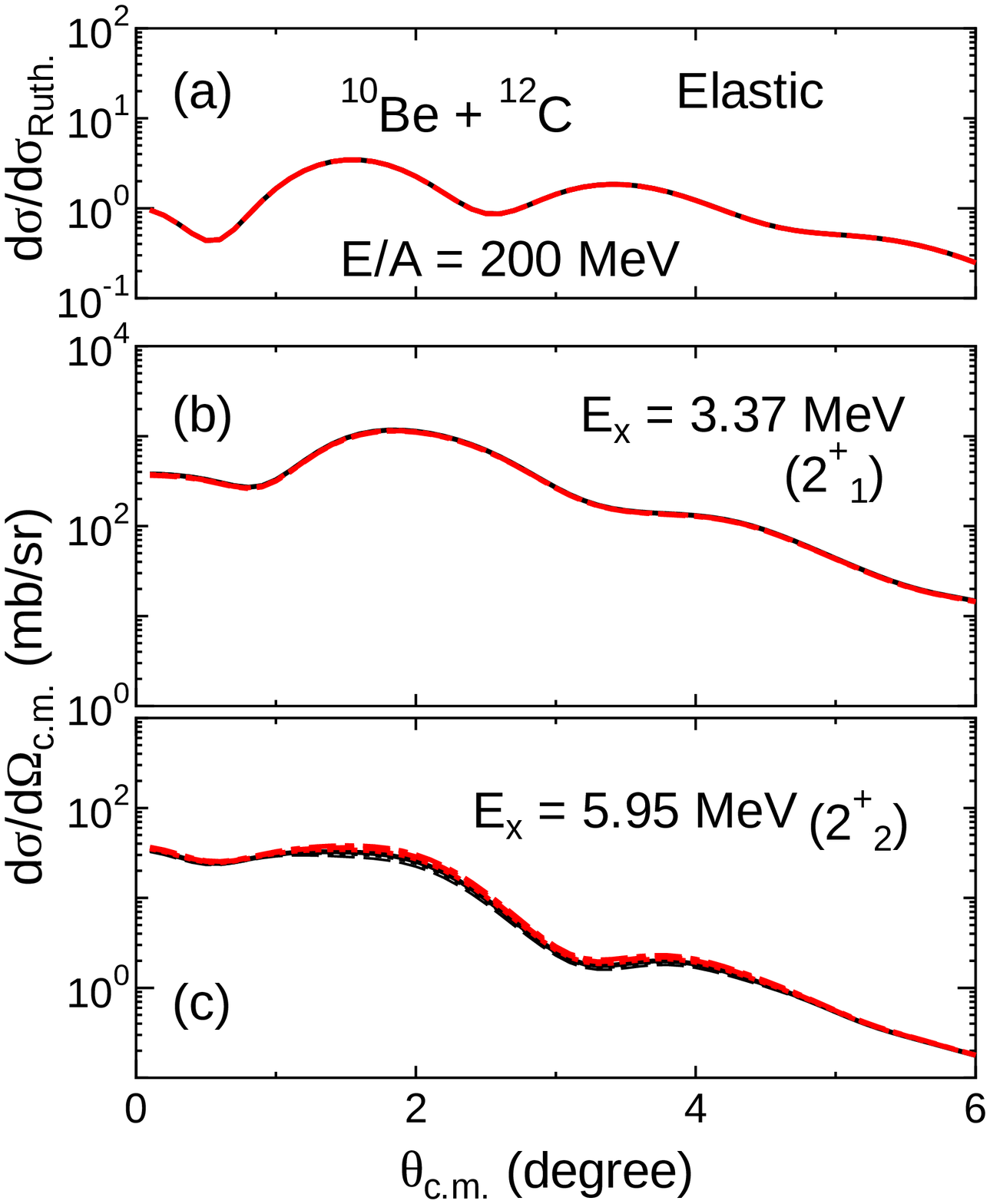}
\caption{\label{fig:pbar2-200mev}
Same as Fig.~\ref{fig:pbar-200mev} but for the $^{12}$C target.}
\end{figure}

Figures~\ref{fig:pbar-59.4mev} and \ref{fig:pbar-200mev} show the calculated elastic and inelastic cross sections for the proton target with the experimental data.
The dashed, dotted, bold-solid, bold(red)-dotted, bold(red)-dashed, and bold(red)-dot-dashed curves are obtained with $B=H$ = 0.00, 0.04, 0.08 0.12, 0.16, and 0.20, respectively.
Again, the effect of changing the $B$ and $H$ values is minor in the elastic and inelastic cross sections.
In the low-lying states of $^{10}$Be, the $B$ and $H$ parameters are found to play a minor role.
Nevertheless, ``the bound two neutrons" is unphysical, and these parameters allow the elimination of such conditions; the application of the $B$ and $H$ parameters is to reproduce a realistic physical situation even if the effect is small.
Figures~\ref{fig:pbar2-59.4mev} and \ref{fig:pbar2-200mev} show the calculated elastic and inelastic cross sections for the $^{12}$C target with the experimental data.
Again, we cannot see the effect of changing the $B$ and $H$ parameters in the elastic and inelastic cross sections for the $^{12}$C target.
These results imply that it is difficult to see the 
difference of the $B$ and $H$ parameters in the elastic and inelastic cross sections.

\subsection{Effect of binding energy of the ground state\label{sec:BE}}

In the above subsections, we have only discussed the parameter dependence in nuclear structure calculations.
That is, we did not specifically consider realistic physical points (e.g., binding energies, etc.).
In this subsection, we adjust the $V_{\rm LS}$ and $M$ parameters so that the binding energies of different parameter sets are consistently reproduced to each other as in Ref.~\cite{ITA02}.
In this paper, we first fix the $V_{\rm LS}$ value in the range of 0--4000 MeV and adjust the $M$ parameter to obtain the appropriate binding energy.
On the contrary, if the $M$ parameter is first fixed in the range of 0.52--0.64, the appropriate $V_{\rm LS}$ value could not be obtained for the small $M$ values of 0.52--0.56, when the constraint of reproducing the binding energy is imposed.
The results are summarized in Table~\ref{tab:LSM}.

\begin{table*}[ht]
\caption{Same as Table~\ref{tab:LS} but for $(V_{\rm LS} \ {\rm (MeV)}, M)$ = (0, 0.57), (500, 0.57), (1000, 0.58), (1500,0.59), (2000,0.60), (2500,0.61), (3000,0.63), (3500,0.64), and (4000,0.65).}
\label{tab:LSM}
\begin{tabular}{lccccccccc} \hline \hline
$V_{\rm LS}$ (MeV) & 0 & 500 & 1000 & 1500 & 2000 & 2500 & 3000 & 3500 & 4000  \\
$M$ & 0.57 & 0.57 & 0.58 & 0.59 & 0.60 & 0.61 & 0.63 & 0.64 & 0.65  \\ \hline 
BE (0$^+_1$) (MeV) & -62.79 & -63.15 & -62.79 & -62.77 & -62.97 & -63.34 & -62.33 & -62.91 & -63.61  \\
BE (2$^+_1$) (MeV) & -60.78 & -60.90 & -60.05 & -59.72 & -59.78 & -60.07 & -58.98 & -59.52 & -60.18  \\
BE (2$^+_2$) (MeV) & -59.27 & -59.64 & -59.04 & -58.40 & -57.72 & -55.67 & -54.70 & -54.85 & -55.08  \\ \hline
$r_p$ (0$^+_1$) (fm) & 2.433 & 2.418 & 2.418 & 2.408 & 2.393 & 2.376 & 2.398 & 2.377 & 2.355  \\
$r_p$ (2$^+_1$) (fm) & 2.410 & 2.407 & 2.419 & 2.402 & 2.380 & 2.360 & 2.388 & 2.364 & 2.340  \\
$r_p$ (2$^+_2$) (fm) & 2.388 & 2.374 & 2.397 & 2.438 & 2.474 & 2.504 & 2.575 & 2.603 & 2.635  \\
$r_n$ (0$^+_1$) (fm) & 2.681 & 2.660 & 2.651 & 2.628 & 2.598 & 2.565 & 2.580 & 2.543 & 2.505  \\
$r_n$ (2$^+_1$) (fm) & 2.660 & 2.655 & 2.666 & 2.634 & 2.597 & 2.561 & 2.582 & 2.543 & 2.505  \\
$r_n$ (2$^+_2$) (fm) & 2.639 & 2.621 & 2.645 & 2.695 & 2.737 & 2.770 & 2.846 & 2.872 & 2.898  \\
$r_m$ (0$^+_1$) (fm) & 2.585 & 2.566 & 2.560 & 2.542 & 2.518 & 2.491 & 2.509 & 2.478 & 2.446  \\
$r_m$ (2$^+_1$) (fm) & 2.563 & 2.559 & 2.570 & 2.544 & 2.512 & 2.482 & 2.506 & 2.473 & 2.440  \\
$r_m$ (2$^+_2$) (fm) & 2.542 & 2.525 & 2.549 & 2.595 & 2.635 & 2.667 & 2.741 & 2.767 & 2.796  \\ \hline
$\langle \bm{L} \cdot \bm{S} \rangle$ (0$^+_1$) & -0.00007341 & 0.4572 & 0.7533 & 0.9237 & 1.026 & 1.091 & 1.110 & 1.145 & 1.172  \\
$\langle \bm{L} \cdot \bm{S} \rangle$ (2$^+_1$) & -0.0006060 & 0.1090 & 0.3788 & 0.7384 & 0.9164 & 1.006 & 1.033 & 1.078 & 1.112  \\
$\langle \bm{L} \cdot \bm{S} \rangle$ (2$^+_2$) & -0.001788 & 0.4060 & 0.4069 & 0.1962 & 0.1156 & 0.09737 & 0.09827 & 0.1051 & 0.1307  \\
$\langle \bm{S}^2 \rangle$ (0$^+_1$) & 0.001095 & 0.08877 & 0.2521 & 0.4007 & 0.5181 & 0.6097 & 0.6547 & 0.7160 & 0.7670  \\
$\langle \bm{S}^2 \rangle$ (2$^+_1$) & 0.002028 & 0.02980 & 0.1667 & 0.3979 & 0.5470 & 0.6399 & 0.6839 & 0.7393 & 0.7844  \\
$\langle \bm{S}^2 \rangle$ (2$^+_2$) & 0.001607 & 0.1232 & 0.1971 & 0.1250 & 0.09792 & 0.1055 & 0.1232 & 0.1457 & 0.1756  \\ \hline
$B$(E2: 2$^+_1$ $\to$ 0$^+_1$) ($e^2$ fm$^4$) & 1.817 & 2.594 & 6.401 & 10.67 & 11.82 & 11.80 & 12.84 & 12.27 & 11.60  \\
$B$((E2)$_n$: 2$^+_1$ $\to$ 0$^+_1$) (fm$^4$) & 13.53 & 13.95 & 16.05 & 14.83 & 12.72 & 11.31 & 12.28 & 11.20 & 10.25  \\  
$B$(IS2: 2$^+_1$ $\to$ 0$^+_1$) (fm$^4$) & 25.26 & 28.58 & 42.73 & 50.67 & 49.06 & 46.20 & 50.24 & 46.93 & 43.66 \\  
$B$(IV2: 2$^+_1$ $\to$ 0$^+_1$) (fm$^4$) & 5.429 & 4.514 & 2.181 & 0.3407 & 0.01631 & 0.005189 & 0.006227 & 0.02422 & 0.04227 \\  
$B$(E2: 2$^+_2$ $\to$ 0$^+_1$) ($e^2$ fm$^4$) & 11.77 & 10.32 & 6.559 & 2.117 & 0.6455 & 0.2236 & 0.1388 & 0.03778 & 0.0009486  \\
$B$((E2)$_n$: 2$^+_2$ $\to$ 0$^+_1$) (fm$^4$) & 3.594 & 2.303 & 0.1438 & 1.022 & 2.640 & 3.439 & 3.758 & 4.007 & 4.169  \\  
$B$(IS2: 2$^+_2$ $\to$ 0$^+_1$) (fm$^4$) & 28.38 & 22.38 & 8.646 & 0.1970 & 0.6747 & 1.909 & 2.452 & 3.267 & 4.044  \\
$B$(IV2: 2$^+_2$ $\to$ 0$^+_1$) (fm$^4$) & 2.358 & 2.874 & 4.760 & 6.081 & 5.897 & 5.417 & 5.341 & 4.823 & 4.295  \\  
$B$(E2: 2$^+_2$ $\to$ 2$^+_1$) ($e^2$ fm$^4$) & 8.154 & 11.22 & 18.98 & 11.34 & 4.390 & 1.851 & 1.408 & 0.6334 & 0.2230  \\
$B$((E2)$_n$: 2$^+_2$ $\to$ 2$^+_1$) (fm$^4$) & 0.002228 & 0.9923 & 13.58 & 22.47 & 17.21 & 12.55 & 11.63 & 8.580 & 6.004  \\
$B$(IS2: 2$^+_2$ $\to$ 2$^+_1$) (fm$^4$) & 8.426 & 18.89 & 64.66 & 65.73 & 38.98 & 24.05 & 21.13 & 13.88 & 8.542  \\
$B$(IV2: 2$^+_2$ $\to$ 2$^+_1$) (fm$^4$) & 7.886 & 5.539 & 0.4507 & 1.886 & 4.213 & 4.764 & 4.944 & 4.551 & 3.913  \\\hline \hline
  \end{tabular}
\end{table*}

Table~\ref{tab:LSM} shows the calculated values of binding energies, radii, expectation values, and quadrupole transition strengths with the $V_{\rm LS}$ and $M$ values obtained with parameters adjusting the binding energy of the ground state.
The binding energies of the ground state are obtained within the deviation of 1 MeV from the set of $V_{\rm LS}$ = 2000 MeV and $M$ = 0.60.
The parameter set is also shown in Table~\ref{tab:LSM}.
On the other hand, the excitation energies of the 2$^+_1$ and 2$^+_2$ states still depend on the set of the $V_{\rm LS}$ and $M$ values.
Here, we notice that the effect of the $V_{\rm LS}$ parameter is dominant for the excitation energy.
By adjusting the binding energy of the ground state, the nuclear sizes of the ground state are obtained almost consistently.
The expectation values of $\langle \bm{L} \cdot \bm{S} \rangle$ and $\langle \bm{S}^2 \rangle$ are drastically dependent on the parameter set; they are strongly dependent on the $V_{\rm LS}$ parameter.
Namely, the property of the $^{10}$Be nucleus strongly depends on the $V_{\rm LS}$ parameter even if we adjust the binding energies of the ground state.

For the proton part of the transition strength from the 2$^+_1$ state to the 0$^+_1$ state, the difference of the $V_{\rm LS}$ and $M$ values
is not clearly seen.
On the other hand, the neutron part shows that the change of the $V_{\rm LS}$ parameter has a significant effect.
For the transition strength from the 2$^+_2$ state to the 2$^+_1$ state, the contribution of the $V_{\rm LS}$ parameter is dominant.
Namely, the obtained nuclear structure information is strongly affected by the $V_{\rm LS}$ parameter even if the energies of the ground state are consistently by simultaneously changing the $V_{\rm LS}$ and $M$ parameters.

\begin{figure}[ht]
\centering
\includegraphics[width=6.4cm]{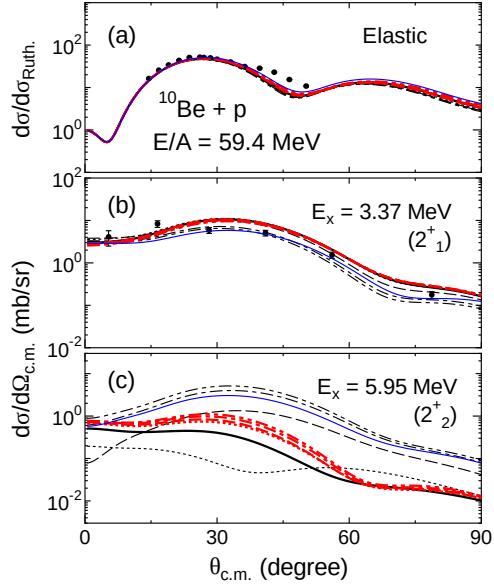}
\caption{\label{fig:LSMdi-59.4mev}
Same as Fig.~\ref{fig:LS-59.4mev} but with the modification of $M$.
The detail is written in the text.
In addition, the (blue) solid curve is  the result obtained with the dineutron basis. The detail of the solid curve is introduced in the next section.}
\end{figure}

\begin{figure}[ht]
\centering
\includegraphics[width=6.4cm]{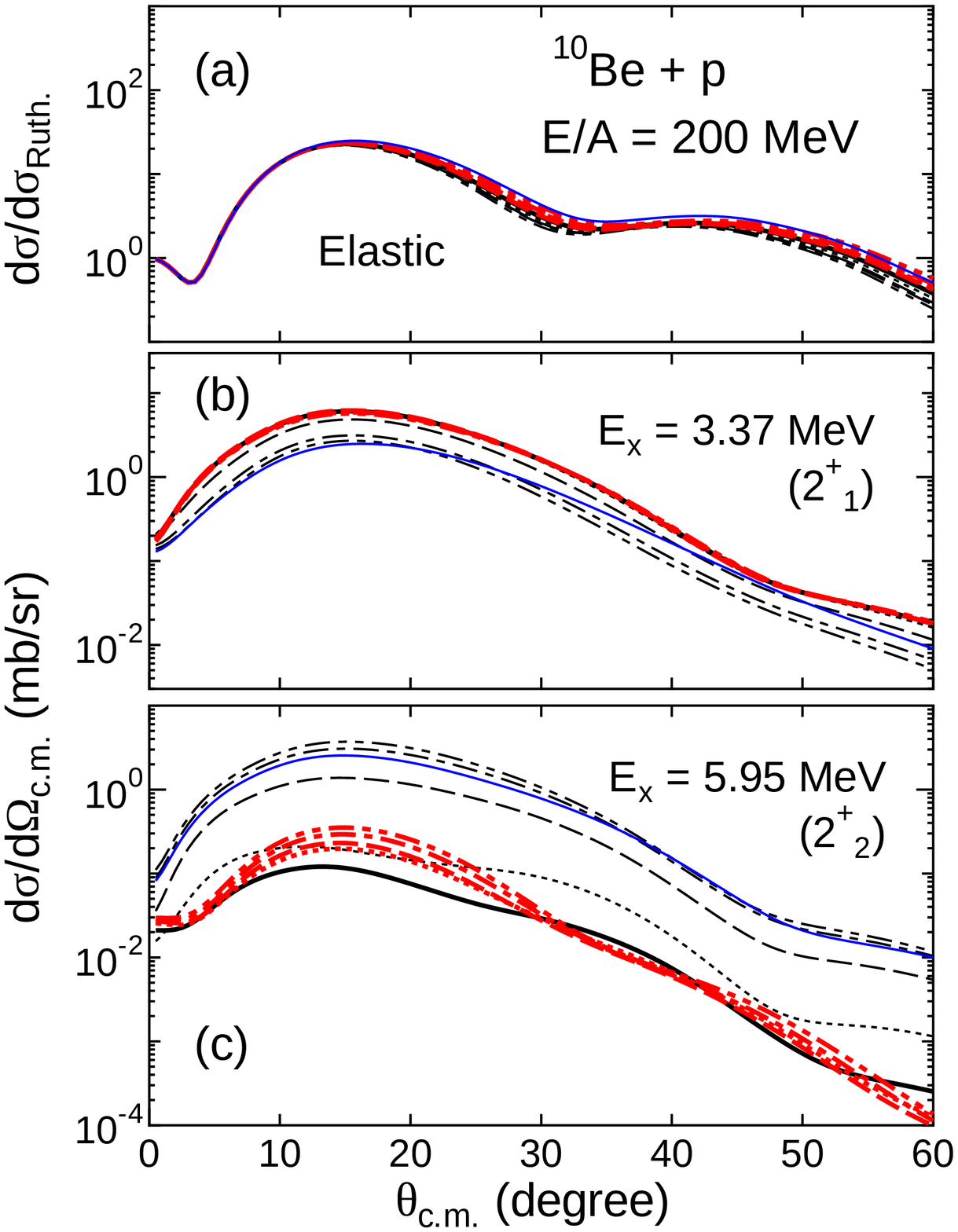}
\caption{\label{fig:LSMdi-200mev}
Same as Fig.~\ref{fig:LSMdi-59.4mev} but at $E/A =$ 200 MeV.}
\end{figure}

\begin{figure}[ht]
\centering
\includegraphics[width=6.4cm]{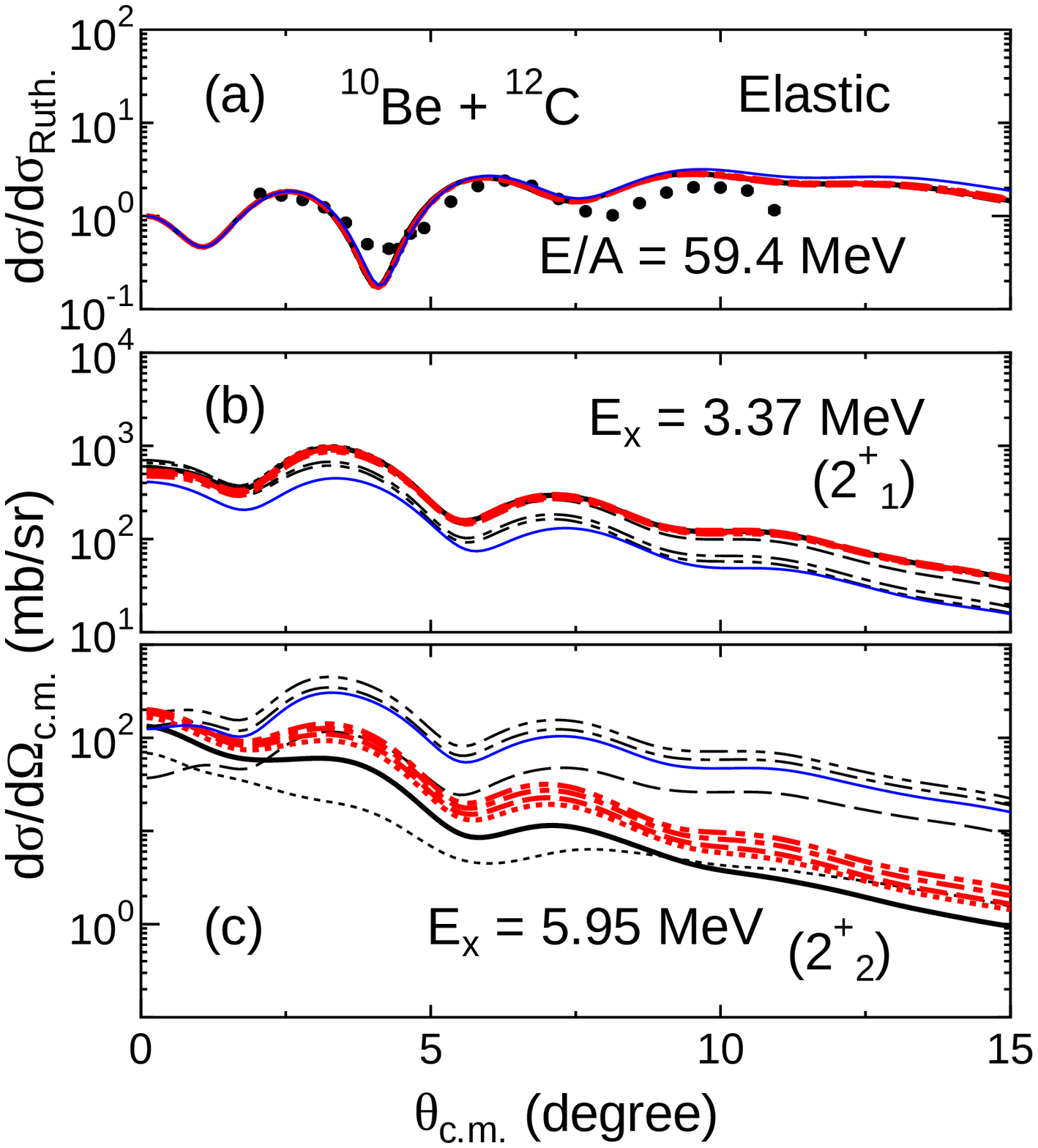}
\caption{\label{fig:LSMdi2-59.4mev}
Same as Fig.~\ref{fig:LSMdi-59.4mev} but for the $^{12}$C target.}
\end{figure}

\begin{figure}[ht]
\centering
\includegraphics[width=6.4cm]{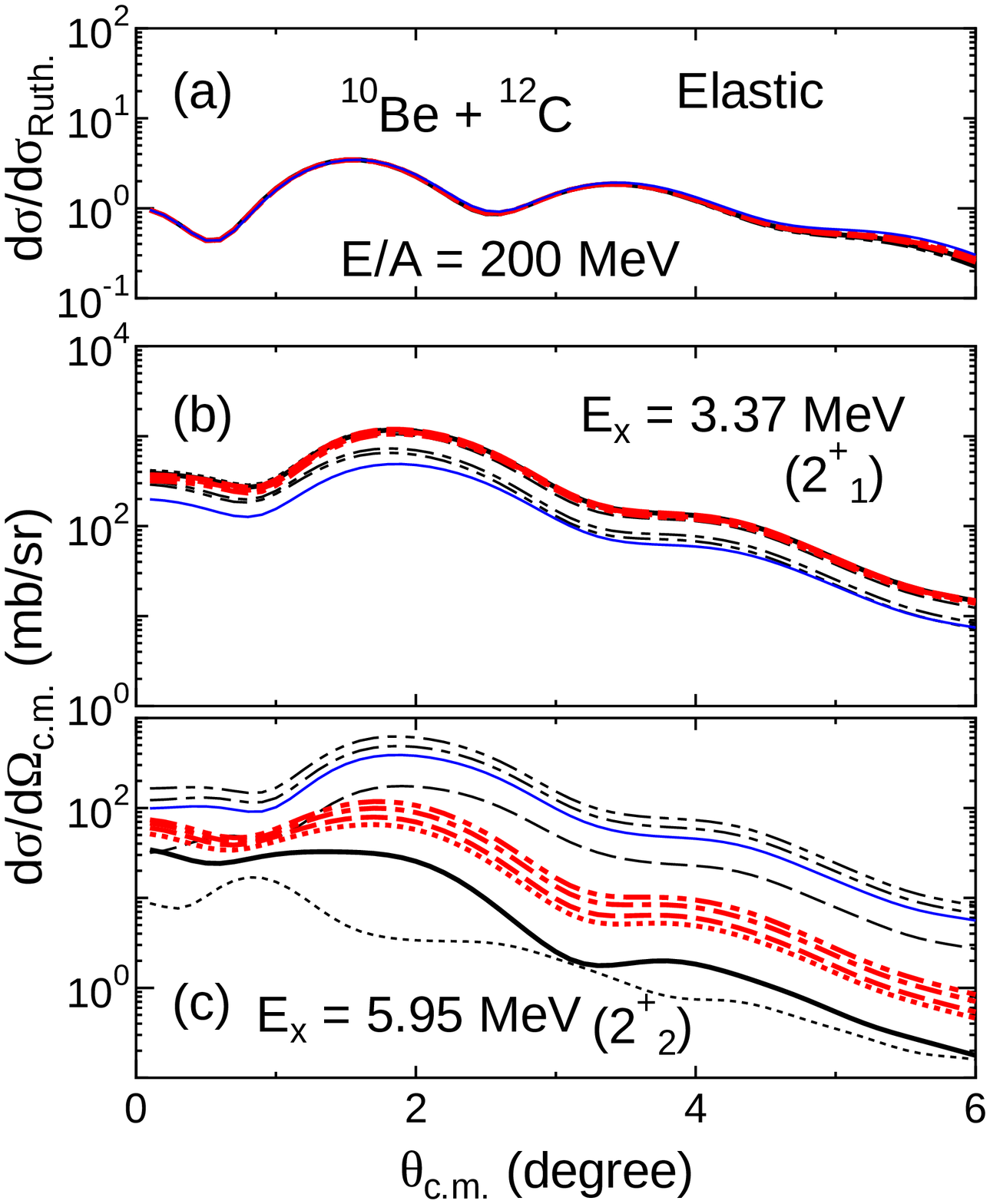}
\caption{\label{fig:LSMdi2-200mev}
Same as Fig.~\ref{fig:LSMdi-200mev} but for the $^{12}$C target.}
\end{figure}

Figures~\ref{fig:LSMdi-59.4mev} and \ref{fig:LSMdi-200mev} show the calculated elastic and inelastic cross sections for the proton target with the experimental data.
The two-dot-dashed, dot-dashed, dashed, dotted, bold-solid, bold(red)-dotted, bold(red)-dashed, bold(red)-dot-dashed, and bold(red)-two-dot-dashed curves are obtained with $(V_{\rm LS} \ {\rm (MeV)}, M)$ = (0, 0.57), (500, 0.57), (1000, 0.58), (1500,0.59), (2000, 0.60), (2500, 0.61), (3000, 0.63), (3500, 0.64), and (4000, 0.65), respectively.
The meaning of the (blue) solid curves will be explained in the next subsection.
As mentioned before, adjusting the binding energy results in a similar nuclear size.
We found that this gives similar elastic cross sections.
The inelastic cross section for the 2$^+_1$ state is slightly affected by the adopted set of the $V_{\rm LS}$ and $M$ parameters.
On the other hand, the drastic change can be seen in the inelastic cross section for the 2$^+_2$ state.
In addition, the results are similar to Figs.~\ref{fig:LS-59.4mev} and \ref{fig:LS-200mev} which are obtained by changing only the $V_{\rm LS}$ parameter.

Figures~\ref{fig:LSMdi2-59.4mev} and \ref{fig:LSMdi2-200mev} show the calculated elastic and inelastic cross sections for the $^{12}$C target with the experimental data.
Again, we can see the consistent results of the elastic cross sections owing to a similar nuclear size.
In addition, we can see the drastic change of the inelastic cross section for the 2$^+_2$ state.
These results are also similar to Figs.~\ref{fig:LS2-59.4mev} and \ref{fig:LS2-200mev}.
Namely, the obtained results support the strong contribution of the $V_{\rm LS}$ parameter even if the energy of the ground state is consistently reproduced to each other by varying the $V_{\rm LS}$ and $M$ parameters simultaneously.

\subsection{Reconfirmation of behavior of inelastic cross section dependent on development and breaking of dineutron correlation \label{sec:dineutron}}

Next, we reconfirm the characteristic behavior of the inelastic cross section for the 2$^+_2$ state depending on the development and breaking of the dineutron correlation.
We use the wave function constructed in the previous subsection and investigate the dineutron correlation in $^{10}$Be.
To describe the dineutron correlation, we prepare the dineutron cluster as
\begin{equation}
\Psi_{i} = {\cal A}
 \big[ \phi_{\alpha} (\bm{r}_{1-4}, \bm{R}_{1}) \phi_{\alpha} (\bm{r}_{5-8}, \bm{R}_{2}) 
\phi_{2n} (\bm{r}_{9-10}, \bm{R}_{3}) \big]_{i}, \label{eq:dineutron}
\end{equation}
where the positions of the centers of the wave packets for the two valence neutrons are the same.
Here, the valence neutrons have spin up and down.
Therefore, the dineutron cluster fully satisfies the condition of the dineutron correlation ($S=0$ and large spatial overlap).
We fix the $M$ parameter to be 0.54 to reproduce the binding energy of the ground state in the same manner as in the previous subsection.

With the dineutron cluster wave function, the obtained binding energies are -63.46, -61.33, and -60.40 MeV for 0$^+_1$, 2$^+_1$, and 2$^+_2$ states, respectively.
The calculated radii are 2.318, 2.289, 2.256, 2.445, 2.401, 2.386, 2.395, 2.357, and 2.335 fm for $r_p$ (0$^+_1$), $r_p$ (2$^+_1$), $r_p$ (2$^+_2$), $r_n$ (0$^+_1$), $r_n$ (2$^+_1$), $r_n$ (2$^+_2$), $r_m$ (0$^+_1$), $r_m$ (2$^+_1$), and $r_m$ (2$^+_2$), respectively.
The transition strengths are 1.677, 8.898, 18.30, 2.849, 7.940, 1.647, 16.82, 2.354, 7.434, 0.2712, 10.54, and 4.865 ($e^2$)fm$^4$ for $B$(E2: 2$^+_1$ $\to$ 0$^+_1$), $B$((E2)$_n$: 2$^+_1$ $\to$ 0$^+_1$), $B$(IS2: 2$^+_1$ $\to$ 0$^+_1$), $B$(IV2: 2$^+_1$ $\to$ 0$^+_1$), $B$(E2: 2$^+_2$ $\to$ 0$^+_1$), $B$((E2)$_n$: 2$^+_2$ $\to$ 0$^+_1$), $B$(IS2: 2$^+_2$ $\to$ 0$^+_1$), $B$(IV2: 2$^+_2$ $\to$ 0$^+_1$), $B$(E2: 2$^+_2$ $\to$ 2$^+_1$), $B$((E2)$_n$: 2$^+_2$ $\to$ 2$^+_1$), $B$(IS2: 2$^+_2$ $\to$ 2$^+_1$), and $B$(IV2: 2$^+_2$ $\to$ 2$^+_1$), respectively.
The excitation energy is similar to the case with the weak $V_{\rm LS}$ parameter rather than the large one.
The nuclear size obtained with the pure dineutron cluster is smaller than all results in this paper.
It is obvious that the expectation values of $\langle \bm{L} \cdot \bm{S} \rangle$ and $\langle \bm{S}^2 \rangle$ are zero.
The transition strengths are also similar to the weak $V_{\rm LS}$ cases rather than the large one while the nuclear size is small.
We note that the $B$(E2: 2$^+_2$ $\to$ 2$^+_1$) value is smaller than the results obtained with $V_{\rm LS}$ = 0--1000 MeV shown in Table~\ref{tab:LS}.
This small $B$(E2: 2$^+_2$ $\to$ 2$^+_1$) value implies the development of the $\alpha$ + $\alpha$ + dineutron cluster picture.
Again, the coupling pattern of the angular momentum among the two $\alpha$ clusters and the dineutron cluster is different in the 2$^+_1$ and 2$^+_2$ states.
Concretely, when the small $V_{\rm LS}$ value is adopted, it is considered that the two $\alpha$ clusters with the relative angular momentum $L=0$ are synthesized with the dineutron around them with $L=2$ in the 2$^+_1$ state. 
On the other hand, the two $\alpha$ clusters with the relative angular momentum $L=2$ are synthesized with the dineutron moving around them with $L=0$ in the 2$^+_2$ state.
The decrease of the $B$(E2: 2$^+_2$ $\to$ 2$^+_1$) value and the increase of the $B$(IV2: 2$^+_2$ $\to$ 2$^+_1$) value corroborate the situation.
In addition, the result of the pure dineutron cluster implies that the developed dineutron cluster can be generated by employing the weak $V_{\rm LS}$ parameter.

Figures~\ref{fig:LSMdi-59.4mev}, \ref{fig:LSMdi-200mev}, \ref{fig:LSMdi2-59.4mev}, and \ref{fig:LSMdi2-200mev} show the result of the pure dineutron cluster state with the (blue) solid curves.
The effect of the nuclear size is not clearly seen on the elastic cross section.
The results of the pure dineutron cluster state are consistent with those obtained with the weak $V_{\rm LS}$ value, especially for the inelastic cross section of the 2$^+_2$ state.
Again, the development of the dineutron structure by weak $V_{\rm LS}$ parameter is supported.

\begin{table*}[ht]
\caption{The components of dineutron configurations $C^{\rm dineutron}(I^\pi_i)$ defined in Eq. (\ref{eq:dineutron2}).
The components are obtained with the wave functions constructed in Secs.~\ref{sec:LS}, \ref{sec:M}, and \ref{sec:BE}.}
\label{tab:ovlap}
\begin{tabular}{lccccccccc} \hline \hline
$V_{\rm LS}$ (MeV) & 0 & 500 & 1000 & 1500 & 2000 & 2500 & 3000 & 3500 & 4000  \\
\multicolumn{10}{c}{$M$=0.60} \\ \hline
$C^{\rm dineutron}$ (0$^+_1$) & 0.8516 & 0.8162 & 0.7562 & 0.6952 & 0.6424 & 0.5976 & 0.5577 & 0.5201 & 0.4839  \\
$C^{\rm dineutron}$ (2$^+_1$) & 0.8125 & 0.7955 & 0.7555 & 0.6634 & 0.5763 & 0.5153 & 0.4665 & 0.4252 & 0.3903  \\
$C^{\rm dineutron}$ (2$^+_2$) & 0.8345 & 0.7740 & 0.7074 & 0.7067 & 0.7129 & 0.6994 & 0.6773 & 0.6512 & 0.6229  \\ \hline \hline
  \end{tabular}
\\
\vspace{3mm}
\begin{tabular}{lccccccccc} \hline \hline
\multicolumn{8}{c}{$V_{\rm LS}$ = 2000 MeV}  \\
$M$ & 0.52 & 0.54 & 0.56 & 0.58 & 0.60 & 0.62 & 0.64 \\ \hline 
$C^{\rm dineutron}$ (0$^+_1$) & 0.5764 & 0.6003 & 0.6197 & 0.6338 & 0.6424 & 0.6465 & 0.6476  \\
$C^{\rm dineutron}$ (2$^+_1$) & 0.4951 & 0.5187 & 0.5401 & 0.5589 & 0.5763 & 0.5936 & 0.6095  \\
$C^{\rm dineutron}$ (2$^+_2$) & 0.7811 & 0.7761 & 0.7632 & 0.7421 & 0.7129 & 0.6777 & 0.6432  \\ \hline \hline
  \end{tabular}
\\
\vspace{3mm}
\begin{tabular}{lccccccccc} \hline \hline
$V_{\rm LS}$ (MeV) & 0 & 500 & 1000 & 1500 & 2000 & 2500 & 3000 & 3500 & 4000  \\
$M$ & 0.57 & 0.57 & 0.58 & 0.59 & 0.60 & 0.61 & 0.63 & 0.64 & 0.65  \\ \hline 
$C^{\rm dineutron}$ (0$^+_1$) & 0.8794 & 0.8376 & 0.7603 & 0.6939 & 0.6424 & 0.6020 & 0.5724 & 0.5444 & 0.5195  \\
$C^{\rm dineutron}$ (2$^+_1$) & 0.8524 & 0.8336 & 0.7628 & 0.6540 & 0.5763 & 0.5234 & 0.4906 & 0.4574 & 0.4294  \\
$C^{\rm dineutron}$ (2$^+_2$) & 0.8762 & 0.8016 & 0.7316 & 0.7276 & 0.7129 & 0.6868 & 0.6461 & 0.6192 & 0.5940  \\ \hline \hline
  \end{tabular}
\end{table*}

To confirm the development of the dineutron structure by weak $V_{\rm LS}$ parameter, we calculate the component of dineutron cluster, while this component is sensitive to basis states dependent on the random number.
Here, we confirmed that the dependence of the basis states does not affect our conclusion.
The component of the dineutron cluster for each state is listed in Table~\ref{tab:ovlap}.
Here the dineutron cluster component ($C^{\rm dineutron}(I^\pi_i)$) for the state $\Phi (I^\pi_i)$ is defined as
\begin{equation}
C^{\rm dineutron}(I^\pi_i) = \sum_k \langle \Phi (I^\pi_i) | \psi_k^{\rm dineutron} \rangle \langle \psi_k^{\rm dineutron} | \Phi (I^\pi_i) \rangle, \label{eq:dineutron2}
\end{equation}
where $\psi_k^{\rm dineutron}$ is the $k$-th orthonormal state obtained by diagonalizing the Hamiltonian matrix only within the basis states of dineutron clusters introduced in Eq.~(\ref{eq:dineutron}).
To clarify on which parameters the dineutron cluster component depends, we show in Table~\ref{tab:ovlap} not only the realistic cases but also cases where only the $V_{\rm LS}$ parameter or the $M$ parameter is changed.
The 0$^+_1$, 2$^+_1$, and 2$^+_2$ states with the weak $V_{\rm LS}$ value have strong overlaps with the pure dineutron cluster state.
With the increase of the $V_{\rm LS}$ value, the dineutron component becomes small.
On the other hand, the components of dineutron configurations have a minor dependence on the $M$ parameter for the 0$^+_1$ and 2$^+_1$ states.
The dependence of the $M$ parameter for the 2$^+_2$ state can be seen, however, the magnitude is smaller than the dependence of the $V_{\rm LS}$ parameter.
This supports that the weak $V_{\rm LS}$ parameter gives the development of the dineutron cluster state in the $^{10}$Be nucleus.

\section{Summary}
\label{sec:summary}

We summarised the relation between the nuclear structure and reaction with the $^{10}$Be nucleus by systematic analysis.
The $^{10}$Be nucleus is constructed by changing the parameters of the effective $NN$ interaction in the microscopic cluster model.
The elastic and inelastic cross sections for the proton and $^{12}$C targets at $E/A=$ 59.4 and 200 MeV were calculated with 
the diagonal and transition densities obtained with various interaction parameters in the MCC calculation.
The calculated elastic and inelastic cross sections well reproduce the experimental data.

In the construction of the $^{10}$Be nucleus, the strength of the spin-orbit potential ($V_{\rm LS}$), the nucleon-nucleon exchange effect ($M$), and the strength of the proton-proton (neutron-neutron) and proton-neutron interactions ($B$ and $H$) are modified.
The $V_{\rm LS}$ parameter has a strong effect on the nuclear structure.
The binding energy, nuclear size, expectation value, and transition strength are drastically changed by the $V_{\rm LS}$ value.
The elastic and inelastic cross sections are also dependent on the $V_{\rm LS}$ parameter.
Especially, the value has an important role in the inelastic cross section of the 2$^+_2$ state.
It is found that the $V_{\rm LS}$ parameter has an important role in the development of the dineutron correlation.
The difference of the IV component is slightly seen in the comparison of the proton and $^{12}$C targets.
The $M$ parameter also has an effect on the nuclear structure information.
The calculated cross sections are also affected by this difference.
The multistep reaction effect is also visible at $E/A =$ 59.4 MeV because the transition strength from the 2$^+_2$ state to the 2$^+_1$ state increases with the $M$ parameter.
On the other hand, we see a clear result without the multistep reaction effect at $E/A =$ 200 MeV.
Comparing the results at 200 MeV, the effect of the $V_{\rm LS}$ parameter on the inelastic cross section of the 2$^+_2$ state is 
found to be larger than the $M$ parameter.
The $B$ and $H$ parameters give a minor effect on the nuclear structure and reaction.
It is the $V_{\rm LS}$ parameter that changes the inelastic cross section of the 2$^+_2$ state most drastically.
Since the degree of development of the dineutron correlation is highly dependent on the strength of the spin-orbit interaction, the structure of $^{10}$Be changes significantly and the cross section also changes with the $V_{\rm LS}$ parameter.

Next, we adjusted the $V_{\rm LS}$ and $M$ parameters simultaneously to reproduce the binding energy of the ground state.
We prepared different sets of these parameters and compared the obtained structure and the effect on the reaction.
As a result, the property of the nuclear structure is found to strongly depend on the $V_{\rm LS}$ parameter rather than the $M$ parameter.
The calculated cross sections are also dependent on the $V_{\rm LS}$ parameter.
In addition, we prepared the pure dineutron cluster wave function to investigate the dineutron component in the $^{10}$Be nucleus by adjusting the $V_{\rm LS}$ and $M$ parameters.
It is found that the $^{10}$Be nucleus with the weak $V_{\rm LS}$ value has a large overlap to the pure dineutron cluster model.
Therefore, we again conclude that the development and breaking of the dineutron correlation in $^{10}$Be is sensitive to changes in the spin-orbit contribution, thereby resulting in drastic changes in the inelastic scattering for the 2$^+_2$ state.

\vspace{2mm}
\acknowledgments
This work was supported by Japan Society for the Promotion of Science (JSPS) KAKENHI Grant Numbers JP20K03944, JP21K03543, and JP22K03618.
This work was supported by the collaborative research program at Hokkaido University.

\end{document}